\documentclass[12pt]{article} 
\pdfoutput=1
\usepackage{amssymb,graphicx}
\usepackage{epstopdf}
\usepackage{amsmath,amsfonts}
\usepackage{epsfig} 
\usepackage{graphicx,graphics}
\usepackage[dvipsnames]{xcolor}
\usepackage{hyperref}
\newcommand{\CLns}{{\tt ${\mathcal C}$osmo${\mathcal L}$attice}}

\begin{document} 

\title{\bf Numerical analysis of melting domain walls and their gravitational waves}
\author{I.~Dankovsky$^{a,b}$, S.~Ramazanov$^c$, E.~Babichev$^d$, D.~Gorbunov$^{b, e}$, A.~Vikman$^f$\\
\small{\em $^a$Faculty of Physics, MSU, 119991 Moscow, Russia}\\
\small{\em $^b$Institute for Nuclear Research of the Russian Academy of Sciences, 117312 Moscow, Russia}\\
\small{\em  $^c$Institute for Theoretical and Mathematical Physics, MSU, 119991 Moscow, Russia}\\
\small{\em $^d$Universit\'e Paris-Saclay, CNRS/IN2P3, IJCLab, 91405 Orsay, France}\\
\small{\em $^e$Moscow Institute of Physics and Technology, 141700 Dolgoprudny, Russia}\\
\small{\em $^f$CEICO, Institute of Physics of the Czech Academy of Sciences (FZU),}\\ 
\small{\em Na Slovance 1999/2, 182 00 Prague 8, Czechia}
}
 
 \date{}

{\let\newpage\relax\maketitle}

\begin{abstract}

We study domain walls (DWs) arising in field theories where $Z_2$-symmetry is spontaneously broken by a scalar expectation value decreasing 
proportionally to the Universe temperature. The energy density of such melting DWs redshifts sufficiently fast not to overclose the Universe. 
For the first time, evolution of melting DWs and the resulting gravitational waves (GWs) is investigated numerically using lattice simulations. We show that formation of closed melting DWs during radiation domination is much more efficient compared to the scenario with constant tension DWs. This suggests that it can be the main mechanism responsible for reaching the scaling regime similarly to the case of cosmic strings. However, the scaling behaviour of 
melting DWs is observed, provided only that the initial scalar field fluctuations are not very large. Otherwise, simulations reveal violation of the scaling law, potentially of the non-physical origin. The spectrum of GWs emitted by melting DWs is also significantly different from that of constant tension DWs. Whether the system has reached scaling or not, the numerical study reveals a GW spectrum described in the infrared by the spectral index $n \approx 1.6$ followed by the causality tail. We attribute the difference from the value $n=2$ predicted in our previous studies to a finite lifetime of the DW network. Notably, the updated index is still in excellent agreement with the recent findings by pulsar timing arrays, which confirms that melting DWs can be responsible for the observed (GW) signal. We also point out that results for evolution of melting DWs in the radiation-dominated Universe are applicable to constant tension DW evolution in the flat spacetime. 

\end{abstract}

\section{Introduction}
\label{sec:intro}

Topological defects like domain walls (DWs)~\cite{Zeldovich:1974uw} and cosmic strings~\cite{Kibble:1976sj} are promising sources of primordial gravitational waves (GWs) in the early Universe~\cite{Vilenkin, Caprini:2018mtu, Hiramatsu:2013qaa}. Searching for GWs generated by dynamics of these topological objects gives an opportunity to probe symmetry breaking patterns associated with otherwise inaccessible energy scales of new physics, or study beyond the Standard Model scenarios in a very feebly coupled regime, 
as is the case of axion-like particles. Interestingly, the correlated signal found recently in pulsar timing measurements\,\cite{NANOGrav:2023gor, NANOGrav:2023hfp, Agazie:2024kdi, EPTA:2023fyk, EPTA:2023xxk, Reardon:2023gzh, Xu:2023wog, InternationalPulsarTimingArray:2023mzf} can be interpreted in terms of GWs from DWs~\cite{NANOGrav:2023hvm}. Furthermore, in Ref.~\cite{Babichev:2023pbf} it has been concluded using theoretical arguments that the subclass of DWs characterised by a time-decreasing tension~\cite{Vilenkin:1981zs, Ramazanov:2021eya, Babichev:2021uvl} (mass per unit area) provides a particularly good fit to the pulsar timing array (PTA) data. Here we support this statement by numerically simulating melting DW network with the help of \CLns~\cite{Figueroa:2020rrl, Figueroa:2021yhd, Figueroa:2023xmq}.

Melting walls are predicted in simple renormalisable particle physics scenarios~\cite{Ramazanov:2021eya, Babichev:2021uvl}, which also provide a promising dark matter candidate. Like conventional DWs, melting ones arise due to spontaneous breaking of a discrete symmetry caused by a non-zero expectation value $\eta$ of a scalar field. However, in the case of melting DWs the symmetry breaking is achieved without introducing a new energy scale: the expectation value is linked to the Universe temperature\footnote{Compared to standard ``melting" in the solid-liquid system, here the symmetry gets broken with {\it increase} of the temperature.} $\eta \propto T$. Consequently, the wall tension $\sigma_{wall} \propto T^3$ can reach large values in the early Universe allowing for an efficient GW production. The energy density of these melting DWs rapidly decreases, 
$\rho_{wall} \sim \sigma_{wall} H \propto 1/a^5$, during radiation domination, so that they get diluted in the surrounding plasma. This is in contrast to the case of conventional DWs, which give a growing contribution to the Universe energy budget. To keep this contribution under control, either 
a severe constraint on the tension should be applied or some mechanism of DW annihilation, e.g., by the explicit breaking of a discrete symmetry~\cite{Zeldovich:1974uw, Gelmini:1988sf} in the early Universe, must be implemented\footnote{See Refs.~\cite{Kibble:1982dd, Coulson:1995nv, Larsson:1996sp, Blasi:2022woz, Blasi:2023rqi} for other solutions of DW problem or the situations, where the problem does not arise in the first place.}. No such a mechanism is needed in the case of melting DWs, where they never dominate in the expanding Universe and finally may disappear in the inverse phase transition~\cite{Weinberg:1974hy}, where the $Z_2$-symmetry gets restored~\cite{Vilenkin:1981zs, Ramazanov:2021eya}. 

As it is mentioned above, recent measurements of the stochastic GW background by various PTAs are in a very good agreement with theoretical expectations for GWs emitted by melting DWs. 
In particular, the spectral shape of GWs is fitted by the power-law $\Omega_{gw} (f) \propto f^{n}$ with $n=1.8 \pm 0.6$. 
This is to be compared with the prediction following from melting DWs $n=2$ in the close-to-maximum infrared frequency range~\cite{Babichev:2023pbf, Babichev:2021uvl}. Nevertheless, this prediction has been obtained 
using some assumptions, in particular, that melting DWs enter the scaling regime. This is the reason, why it is necessary to verify the theoretical expectations 
of Refs.\,\cite{Babichev:2023pbf, Ramazanov:2021eya,Babichev:2021uvl} with numerical simulations.

Notably, melting DWs are much more suited to lattice simulations compared to constant tension DWs. Characterised by a constant wall width (therefore, we will use the term constant width DWs interchangeably), the latter eventually become thinner than the lattice spacing, which grows as the Universe scale factor, so that one ceases to resolve DWs on a lattice. Melting DWs by contrast have the width growing with the scale factor just like the lattice spacing. 
Hence, if the wall width is initially larger than the lattice spacing, this will remain true at all times. Consequently, the only limitation left when performing simulations with melting DWs is due to a finite box size, which must be kept larger than the simulated Hubble patch. 
As a result, the available (conformal) time span of simulations grows linearly with the grid number $N$, i.e., 
$\Delta \tau \propto N$, which manifests a significant advantage over the case with constant tension walls, where $\Delta \tau \propto \sqrt{N}$.

Let us summarise the main findings of this work. 
First, closed DWs are much more abundantly produced in the case of melting DWs compared to the scenario with constant width walls studied in Ref.~\cite{Dankovsky:2024zvs}. Namely, the ratio of the total area contained in closed walls to the area of a single long wall extending over the simulation box, is about $30\%$ in the melting DW case versus at most a few percent in the conventional scenario. It is likely 
that formation of closed walls eventually being dissolved into a collection of particles is the main channel of reaching the scaling regime by a network of melting walls, akin to the case of cosmic strings~\cite{Matsunami:2019fss, Saurabh:2020pqe, Baeza-Ballesteros:2023say, Baeza-Ballesteros:2024otj}. There is, however, an important qualification, as the scaling of melting walls is attained only for sufficiently small initial scalar field fluctuations. This is the second crucial difference 
with the case of constant width walls, where the scaling solution is attained independently of initial conditions. 
For sufficiently large scalar fluctuations, we observe a gross violation of scaling in the melting DW network. 
Such a violation occurs primarily due to an abundant formation of small substructure all over the broken symmetry stage, while the long wall area still respects the scaling law. At this stage of advancement, we are unable to conclude if this substructure is mainly due to small walls or rather some generic scalar fluctuations misinterpreted as walls. 

The GW spectrum produced by melting DWs exhibits a small, but significant deviation from the prediction $n=2$. As one can see in Fig.~\ref{spectrum_vacuum}, numerical simulations reveal the GW slope $n \approx 1.6$ in the close-to-maximum infrared frequency range. 
Note, however, that the result $n =2$ has been derived in the idealised setup 
where the source of GWs, i.e., melting DWs, 
exists infinitely in the past. In reality, DWs are ``turned on" at some time $\tau_i$, and accounting for $\tau_i$ affects the calculation of GWs 
exactly by decreasing the value of $n$. Remarkably, the result $n \approx 1.6$ holds in the portion of the low frequency part of the spectrum whether the scaling behaviour is obeyed or not, as one can see from the comparison of Figs.~\ref{spectrum_vacuum} and~\ref{spectrum_thermal}. In the high frequency range, GW spectrum exhibits the power law decrease in the scaling case (Fig.~\ref{spectrum_vacuum}) and growth 
in the non-scaling case (Fig.~\ref{spectrum_thermal}). 
We also update the particle physics scenario underlying melting DWs in light of our numerical findings.

Organisation of this paper is similar to that of Ref.~\cite{Dankovsky:2024zvs} discussing evolution of constant tension DWs and resulting GW emission. This can facilitate the comparison between two scenarios, with constant tension and melting DWs. See also earlier works~\cite{Hiramatsu:2013qaa, Kitajima:2023cek, Ferreira:2023jbu, Kitajima:2023kzu} on phenomenological consequences of constant tension DWs obtained with high resolution simulations. In Section~\ref{sec:basic}, we briefly review the main properties of melting DWs. We perform theoretical estimation of GW spectrum produced by melting DWs in Section~\ref{gw:theory}. After adjusting the system for lattice simulations in Section~\ref{sec:optimisation}, we study numerically DW evolution in Section~\ref{sec:gwnumerics}. Results of lattice simulations for the GW spectrum are discussed in Section~\ref{sec:gwnumerics}. In Section~\ref{sec:applications}, we review a concrete particle physics scenario proposed in Ref.~\cite{Ramazanov:2021eya} leading to melting DWs. 
There we also discuss the possibility that a realisation of this scenario results in the observed PTA signal. Using numerical results on GW spectrum obtained in Section~\ref{sec:gwnumerics}, we determine the couplings, for which the model predictions match observations. We conclude in Section~\ref{sec:discussions}.

\section{Basic properties of melting domain walls}
\label{sec:basic}
We consider the model of a canonical scalar field $\chi$ with the double well potential which allows for spontaneous breaking of $Z_2$-symmetry:
\begin{equation}
\label{Lagrange}
{\cal L}=\frac{1}{2} (\partial_{\mu} \chi)^2-\frac{1}{4} \lambda_{\chi}  \left(\chi^2-\eta^2 (T)\right)^2 \; .
\end{equation}
The difference compared to the standard scenario of constant tension DWs is the expectation value $\eta$, which evolves non-trivially with the Universe temperature $T$. We choose 
\begin{equation}
\label{alphadef}
\eta (T) =\alpha T \; ,
\end{equation}
where $\alpha$ is some model-dependent dimensionless constant, see Section~\ref{sec:applications} for a particular example. The equation of motion following from the Lagrangian~\eqref{Lagrange} is given by
\begin{equation}
\label{scalareq}
\chi''+2 \frac{a'}{a} \chi' -\partial^2_i \chi +\lambda_{\chi}  (\chi^2 a^2 -\eta^2_c) \chi=0 \; .
\end{equation}
Here $a=a(\tau)$ is the scale factor describing the Universe expansion, $\eta_c \equiv \eta (T) \cdot a $ is the comoving expectation value and the prime denotes differentiation with respect to conformal time $\tau$. 
Hereafter, we assume the radiation dominated stage, where the scale factor evolves as $a (\tau) \propto \tau$. Notably, one can dramatically simplify Eq.~\eqref{scalareq} by making the field redefinition: 
\begin{equation}
\chi=\frac{s}{a} \; .
\end{equation}
Then one can rewrite Eq.~\eqref{dimensionless} during radiation domination as follows:
\begin{equation}
\label{Minkowski}
s''-\partial^2_i s +\lambda_{\chi} s (s^2-\eta^2_c) =0 \; .
\end{equation}
Remarkably, the same equation describes evolution of a scalar with a constant expectation value in Minkowski spacetime.

Generically, systems exhibiting spontaneous breaking of discrete symmetries give rise to DWs. One of the main characteristics of DWs is its tension (mass per unit area) given by  
\begin{equation}
\label{meltingtension}
\sigma_{wall}=\frac{2\sqrt{2\lambda_{\chi}} \eta^3}{3} =\frac{2\sqrt{2 \lambda_{\chi}} \alpha^3 }{3}\,\, T^3 \; ,
\end{equation}
where in the second equality we used Eq.~\eqref{alphadef}. In the system of interest the DW tension redshifts as $\sigma_{wall} \propto 1/a^3$. Following Ref.~\cite{Babichev:2021uvl}, we call such topological objects {\it melting DWs}. On the other hand, we refer to conventional DWs predicted in the model with $\eta =\mbox{const}$ as constant tension DWs. As we mentioned above, evolution of melting DWs in the radiation dominated Universe is equivalent to evolution of constant tension walls in Minkowski spacetime. In due course, we make use of this correspondence, when discussing evolution of melting DWs.

It has been shown in Refs.~\cite{Press:1989yh} that in the radiation dominated Universe
the network of constant tension DWs settles to the 
scaling regime after some time. What is more relevant for us is that the same property of constant tension DWs 
holds in Minkowski spacetime~\cite{Garagounis:2002kt}. Thus, according to the discussion above, melting DWs should also obey the scaling law at radiation domination. We check this statement in Section~\ref{sec:dwevolution}, where it is demonstrated that the scaling behaviour is indeed obeyed for some generic initial conditions, though violations of the scaling law (potentially of the systematic origin) are observed for sufficiently large initial fluctuations of the scalar $\chi$. For the time being, we assume that the scaling law is always respected by melting DWs. This means that there is a fixed number of DWs in the horizon volume. Typically, there is one DW stretching throughout the horizon, and hence one can estimate the DW energy density as 
\begin{equation}
\label{area}
\rho_{wall} \simeq 2\sigma_{wall} H \xi \; ,
\end{equation}
where $\xi$ is the so called area parameter. It is defined as
\begin{equation}
\label{scalingdef0}
\xi \equiv \frac{S \, t}{a(t) V} \; ,
\end{equation}
where $S$ is the wall comoving area contained inside the comoving volume $V$ and $t$ denotes cosmological time given by $dt=a(\tau)d\tau$. Crucially, the area parameter is constant in the scaling regime:  
\begin{equation}
\label{scalingscaling}
\xi =\xi_{sc} \approx \mbox{const} \; .
\end{equation}
As we show in what follows, the moment of settling to the scaling regime $\tau_{sc}$ is when the full-fledged DW evolution starts 
(in particular, GW emission is negligible before this moment). 
The Universe temperature $T_{sc}$ at this moment is given by $T_{sc}=T_i \tau_{i}/\tau_{sc}$, where $\tau_i$ is the initial conformal time of the network formation, and $T_i$ is the corresponding Universe temperature. Formation of DWs begins when the scalar $\chi$ with the background value cosmologically set to zero starts to roll towards minima of its spontaneously breaking potential at the conformal time $\tau_i$. Setting the field $\chi$ to zero can be achieved, e.g., during inflation by coupling $\chi$ to the Ricci scalar~\cite{Babichev:2021uvl}. At radiation domination, which is of our main interest, the Ricci scalar is negligible, and we can safely ignore such a coupling. The moment $\tau_i$ is defined by the balance of the field $\chi$ mass and the Hubble friction, i.e., $\sqrt{\lambda_{\chi}} \alpha T_i \sim H(T_i)$. Consequently, we get
\begin{equation}
\label{initscaling}
T_{sc} \approx \frac{3 \alpha \sqrt{\lambda_{\chi}} M_{Pl}}{\sqrt{g_* (T_{sc})}} \cdot \frac{\tau_{i}}{\tau_{sc}} \; ,
\end{equation}
where $g_* (T)$ is the number of relativistic degrees of freedom in the Universe, and $M_{Pl} \approx 2.45 \cdot 10^{18}~\mbox{GeV}$ is the reduced Planck mass. Note that we ignored evolution of $g_* (T)$ from the time 
$\tau_i$ to $\tau_{sc}$. Eventually, the ratio $\tau_i/\tau_{sc}$ is deduced from numerical simulations, so that $T_{sc}$ is fully defined in terms of the model parameters.

Equation~\eqref{area} implies that the energy density of DWs redshifts as $\rho_{wall} \propto 1/a^5$ during radiation domination, i.e., faster than the energy density of the surrounding matter $\rho \propto 1/a^4$. Hence, no problem with overclosing the Universe arises in the melting DW case. In this sense, phenomenological consequences of melting DWs are more robust, because one does not need to introduce a potential or population bias to destroy the network as in the case of constant tension walls. In the latter case the strongest signal comes from the last stage of the network evolution, and the mechanism destroying the DW network may impact on the final form of the GW spectrum, while the details of the phase transition leading to DW production are irrelevant. In the case of melting DWs the situation is quite the opposite. Namely, phenomenological consequences of melting DWs, i.e., the GW signal discussed in the next section, are more sensitive to initial conditions and particularities of settling to the scaling regime. However, this leaves the most crucial predictions of the melting DW scenario intact, as we show in what follows.

\section{Gravitational waves: theoretical expectations}
\label{gw:theory}

DW tension decreasing as $\sigma_{wall} \propto 1/a^3$ leads to drastically different predictions for the GW spectrum compared to constant tension DWs. In accordance with the discussion in the end of the previous section, most of GWs are emitted close to the moment $\tau_i$ of DW formation, or more precisely to the moment of entering the scaling regime $\tau_{sc}$. This introduces some uncertainty when estimating the peak frequency of GW emission. On the other hand, a more conclusive statement can be made about the low frequency part of GW spectrum, sourced by late-time evolution of melting DWs, when they are already in the scaling regime. Indeed, it has been argued in Refs.~\cite{Babichev:2023pbf, Babichev:2021uvl} that the infrared tail of GWs is  described by the spectral index $n=2$. Below we refine this analysis following Ref.~\cite{Ramazanov:2023eau}, and find corrections to $n=2$ due to a finite duration of the source operation, which happens in any realistic model.

The energy density of GWs is given by~\cite{Maggiore} 
\begin{equation}
\label{gwenergy}
\rho_{gw}=\frac{\langle h'_{ij} ({\bf x}, \tau) h'_{ij} ({\bf x}, \tau) \rangle }{32\pi G_N a^2 (\tau)} \; ,
\end{equation}
where $h_{ij}$ is the spin-2 transverse-traceless metric fluctuation (GWs) we are interested in, and $G_N$ is the Newton's constant. Averaging in Eq.~\eqref{gwenergy} is performed over many GW wavelengths/GW periods. The equation of motion describing Fourier modes of GWs is given by 
\begin{equation}
\label{gweq}
h''_{ij} ({\bf k}, \tau)+\frac{2a'}{a} h'_{ij} ({\bf k}, \tau)+k^2 h_{ij} ({\bf k}, \tau) =16\pi G_{N} a^2 \Pi_{ij} ({\bf k}, \tau) \; ,
\end{equation}
where $\Pi_{ij} ({\bf k}, \tau)$ is the Fourier transform of the transverse traceless (TT) part of the source stress-energy tensor, i.e., $\Pi_{ij} ({\bf k}, \tau)=T^{TT}_{ij} ({\bf k}, \tau)/a^2$. Equivalently, one can write
\begin{equation}
\Pi_{ij} ({\bf k}, \tau)=\frac{1}{a^2}\Lambda_{ij, kl} ({\bf \hat{k}}) T_{kl} ({\bf k}, \tau), 
\end{equation}
where $\Lambda_{ij, kl} ({\bf \hat{k}})$ is the Lambda-tensor selecting the TT part of a symmetric tensor and ${\bf \hat{k}}={\bf k}/k$. The solution of Eq.~\eqref{gweq} satisfying initial conditions $h_{ij} ({\bf k}, \tau_i)=0$ and $h'_{ij} ({\bf k}, \tau) |_{\tau_i}=0$, reads 
\begin{equation}
\label{Fourier}
h_{ij} ({\bf k}, \tau) =\frac{16\pi G_{N}}{a(\tau) k} \int^{\tau}_{\tau_i} d\tau' a^3 (\tau')\sin k(\tau-\tau') 
\Pi_{ij} ({\bf k}, \tau') \; .
\end{equation}
This formula is exact during radiation domination, while at later stages it is valid for the modes, which are subhorizon at matter-radiation equality. The wavelengths characteristic for these modes are in the range covering all observational facilities of interest. Therefore, we consider only such modes in what follows. We perform the inverse Fourier transformation, i.e., 
\begin{equation}
h_{ij} ({\bf x}, \tau)= \int d{\bf k} e^{i{\bf kx}} h_{ij} ({\bf k}, \tau) \; ,
\end{equation}
substitute the latter into Eq.~\eqref{gwenergy}, and use Eq.~\eqref{Fourier}. Keeping only sub-horizon modes at the time $\tau$, i.e., $k\tau \gg 1$, 
and averaging out fast oscillating terms, we obtain:
\begin{equation}
\begin{split}
&\rho_{gw} =\frac{4\pi G_N}{a^4(\tau)} \int \int d{\bf k} d{\bf q} e^{i({\bf k}+{\bf q}){\bf x}} \int^{\tau}_{\tau_i} \int^{\tau}_{\tau_i} d\tau' d\tau'' a^3 (\tau') a^3 (\tau'') \times \\ & \times \langle \Pi_{ij} ({\bf k}, \tau') \Pi_{ij}({\bf q}, \tau'') \rangle \cdot 
\cos (k\tau'-q\tau'') \; .
\end{split}
\end{equation}
Then we split the tensor $\Pi_{ij}$ into polarisations:
\begin{equation}
\Pi_{ij} ({\bf k}, \tau) =\sum_{A} \Pi_A ({\bf k}, \tau) \cdot e^{A}_{ij} ({\bf \hat{k}})\; ,
\end{equation}
where the pair of TT polarisation tensors $e^{A}_{ij}$, $A=1, 2$, obeys the normalisation condition $e^{A}_{ij} e^{A'}_{ij}=2 \delta_{AA'}$. Taking into account that the network of melting DWs is statistically a spatially homogeneous and  isotropic source, and that it produces unpolarised GW radiation, one can write~\cite{Caprini:2018mtu}:
\begin{equation}
\label{scalingdef}
\langle \Pi_{A} ({\bf k}, \tau') \Pi_{A'} ({\bf q}, \tau'') \rangle =\delta_{AA'} \delta ({\bf k}+{\bf q}) 
\rho_{wall} (\tau') \rho_{wall} (\tau'') P(k, \tau', \tau'') \; .
\end{equation}
Now let us assume that no efficient GW production takes place before 
the onset of the scaling regime. This allows us to simply replace the initial time with the time when scaling starts in all the formulas above \footnote{Note that the exact $Z_2$-symmetry warrants that formation of DWs occurs during the second order phase transition, and therefore no substantial GW production takes place before the time $\tau_i$.},  \begin{equation}
\tau_i \rightarrow \tau_{sc} \; .
\end{equation}
This will be verified in what follows using numerical simulations.
In the scaling regime, the form of the unequal time power spectrum is fixed to be (see, e.g., Ref.~\cite{Dankovsky:2024zvs})
\begin{equation}
P(k, \tau', \tau'')=(\tau' \tau'')^{3/2} \cdot {\cal P} (k\tau', k\tau'') \; .
\end{equation}
Recall also that the melting DW energy density decreases as $\rho_{wall} \propto 1/a^5$, i.e., $\rho_{wall} \cdot a^5 \approx \mbox{const}$. As a result, the energy density of GWs can be expressed as follows:  
\begin{equation}
\label{verylong}
\begin{split}
&\rho_{gw}=64\pi^2 G_N \rho^2_{wall} (\tau) a^2 (\tau) \tau^4 \int dk k^2 \int^{\tau}_{\tau_{sc}} \int^{\tau}_{\tau_{sc}} \frac{d\tau' d\tau''}{\sqrt{\tau' \tau''}} ~{\cal P} (k\tau', k\tau'') \cdot \cos k(\tau'-\tau'') \; .
\end{split}
\end{equation}
Introducing dimensionless variable $u \equiv k\tau$, we obtain the spectral energy density of GWs in the form:
\begin{equation}
\label{spectrum}
\frac{d\rho_{gw}}{d\ln k} =64 \pi^2 G_N \rho^2_{wall} (\tau)  a^2 (\tau) k^2 \tau^4 \cdot A(u_{sc}, u) \; , 
\end{equation}
where $A(u_{sc}, u)$ is the dimensionless function of dimensionless variables $u\equiv k\tau$ and $u_{sc} \equiv k\tau_{sc}$, 
\begin{equation}
\label{A}
A(u_{sc}, u)=\int^{u}_{u_{sc}} \int^{u}_{u_{sc}} \frac{d u' d u''}{\sqrt{ u' u''}} {\cal P} (u', u'') \cos (u''-u') \; .
\end{equation}
One can further simplify the expression for the spectral energy density by taking the limits $u_{sc} \rightarrow 0$ and $u \rightarrow \infty$ in the double integrals above. These limits imply that we are focusing on the momentum range $2\pi/\tau \ll k \ll 2\pi/\tau_{sc}$. It is convenient to present the final result in terms of fractional spectral energy density of GWs defined as
\begin{equation}
\label{fraction}
\Omega_{gw}  =\frac{1}{\rho_{tot}} \cdot \frac{d\rho_{gw}}{d\ln k}  \; ,
\end{equation}
where $\rho_{tot}$ is the total energy density of the Universe. During the radiation domination era the energy density is $\rho_{tot}=\pi^2 g_* (T) T^4/30$, and from Eqs.~\eqref{spectrum} and~\eqref{fraction} one obtains 
in the momentum range $2\pi/\tau \ll k \ll 2\pi/\tau_{sc}$:
\begin{equation}
\label{spectrumideal}
\Omega_{gw} (k, \tau) \approx \frac{6.8 \cdot 10^3 \, \lambda_{\chi} \, \alpha^6 \, \xi^2_{sc} \, G_N \,  T^2  \, k^2 \tau^2 }{g_* (T)} \cdot A(0, \infty) \; .
\end{equation}
Here we also used Eqs.~\eqref{meltingtension},~\eqref{area},~\eqref{scalingscaling}, and the expression for the Hubble rate $H =\frac{1}{a(\tau) \tau}$. Hence, the spectral shape of GWs in the interesting frequency range is described by the power-law with the exponent $n=2$. This result can be traced back to the fact that the energy density of melting DWs decreases as\footnote{More generically, it has be proven in Ref.~\cite{Ramazanov:2023eau} that a source of GWs in the scaling regime with the energy density behaving as $\rho \propto 1/a^{\beta}$, where $\beta$ is some constant, 
leads to the spectral index $n=2\beta-8$ in a certain range of $\beta$. In particular, one reproduces $n=0$ in the case of cosmic strings described by $\beta=4$, cf. Ref.~\cite{Figueroa:2012kw}.}  $\rho_{wall} \propto 1/a^5$~\cite{Ramazanov:2023eau}. For GWs with wavelengths, which are super-horizon till the final time $\tau$, 
the spectral shape reads 
\begin{equation}
\Omega_{gw} (k) \propto k^3 \qquad k \tau \ll 2\pi\; .
\end{equation}
This is the standard result following from causality considerations~\cite{Caprini:2009fx, Hook:2020phx, Cai:2019cdl, Figueroa:2020lvo} and applicable largely independently of the source nature. We do not repeat its derivation here. 

Note that $\Omega_{gw}$ only very mildly depends on time through the change of the relativistic degrees of freedom number $g_* (T)$. 
This is in a sharp contrast to the case of constant tension DWs, where $\Omega_{gw} (\tau) \propto a^4 (\tau)$ during radiation domination at the times, when the scaling regime is operating~\cite{Hiramatsu:2013qaa}.

The energy density of GWs relies on the unknown coefficient $A$. When DWs dissolve and GW production terminates at some time $\tau_f$, the relative contribution of GWs is given by Eq.\,\eqref{spectrumideal} with $\tau=\tau_f$. Later into radiation domination, the relative contribution of GW remains almost constant, up to corrections due to decoupling of degrees of freedom from the primordial plasma. Below we present the estimate for GW spectrum, and we hide intermediate details of estimating the function $A(u_{sc}, u)$ in the Appendix. Crucially, we do not set to zero and infinity the limits of integration in Eq.~\eqref{A} and estimate the first non-trivial corrections in $u_{sc}$ and in $1/u$. In this way, we properly take into account a finite duration of the source, which yields corrections to the spectral tilt $n=2$. Using Eq.~\eqref{simplethree} there with the constant ${\cal J}$ estimated in Eq.~\eqref{j}, we arrive at the analytical estimate 
of the spectral energy density of GWs:
\begin{equation}
\label{estimate}
\Omega_{gw} (k, \tau) \sim \frac{\lambda_{\chi} \, \alpha^6 \, \xi^2_{sc} \, G_N \, T^2 \, k^2 \tau^2}{g_* (T)} \cdot 
\left(1-\frac{k\tau_{sc}}{2.6 \pi}-0.23 \cdot \left(\frac{2\pi}{k\tau} \right)^{3.5}  \right)\,, \qquad \frac{2\pi}{\tau} \lesssim k \lesssim 
\frac{2\pi}{\tau_{sc}} \; .
\end{equation}
Then the GW spectrum at lower frequencies can be generically parameterised as a power law with slightly running index, 
\begin{equation}
\Omega_{gw} (k, \tau) ={\cal R} (\bar{k}) \cdot \left(\frac{k}{\bar{k}} \right)^{n} \; .
\end{equation}
It is then straightforward to obtain the spectral index from Eq.\,\eqref{estimate}:
\begin{equation}
 n  \sim 2-\frac{(k-\bar{k}) \tau_{sc}}{2.6 \pi \ln k/\bar{k}} -
\frac{0.23}{\ln k/\bar{k}} \cdot \left[\left(\frac{2\pi}{k\tau}\right)^{3.5} - \left(\frac{2\pi}{k\bar{\tau}}\right)^{3.5}\right]\; .
\end{equation}
Here $\bar{k}$ fulfills $2\pi/\tau \ll \bar{k} \ll 2\pi/\tau_{sc}$, but otherwise can be chosen in an arbitrary way. 
Note that we have set $n(\bar{k})=2$. This choice fixes the amplitude 
${\cal R} ({\bar k}, \tau)$ to be 
\begin{equation}
{\cal R} ({\bar k}, \tau) \sim \frac{\lambda_{\chi} \, \alpha^6 \, \xi^2_{sc} \, G_N \, T^2 \, \bar{k}^2 \tau^2}{g_* (T)} \cdot \left(1-\frac{\bar{k} \tau_{sc}}{2.6\pi}-0.23 \left(\frac{2\pi}{\bar{k} \tau} \right)^{3.5} \right) 
\;.
\end{equation}
It is worth mentioning that the NANOGrav collaboration has already initiated the search for the spectral index running~\cite{Agazie:2024kdi}, albeit with a null result at the moment. Hopefully, with increased sensitivity of PTA measurements it will become possible to test the predicted running in the future.

The spectrum exhibits a maximum at a particular frequency $k_{peak}$, which satisfies $k_{peak} \tau_{sc} >2\pi$, as it will become clear in what follows. For such large $k_{peak}\tau_{sc}$, Eq.~\eqref{estimate} is not applicable, and the fractional energy density of GWs at peak is rather estimated as 
\begin{equation}
\label{peakth}
\Omega_{gw, peak} \sim \frac{0.3 \lambda_{\chi} \alpha^6 \xi^2_{sc} G_N T^2_{sc} k^2_{peak} \tau^2_{sc}}{g_* (T_{sc})} \cdot \left(\frac{2\pi}{k_{peak} \tau_{sc}} \right)^{3.5} \; ,
\end{equation}
where we used Eqs.~\eqref{a} and~\eqref{j} from the Appendix. Note that following our analysis in the Appendix, we could also obtain the spectral shape for $k >k_{peak}$. However, for such large $k$ the analogy with constant tension DWs used when deriving Eqs.~\eqref{estimate} and~\eqref{peakth} becomes less trustworthy. The reason is that the small scale structure, to which the ultraviolet part of the spectrum is sensitive, is much more developed in the case of melting DWs compared to constant width ones, as it will become clear in Section~\ref{sec:dwevolution}.

From the observational point of view one is interested in the present day fractional energy density of GWs, which is related to the fractional energy density $\Omega_{gw} (\tau)$ as
\begin{equation}
\label{OmegaOmega}
\Omega_{gw} h^2_0 =1.34 \cdot 10^{-5} \cdot \left(\frac{100}{g_* (T)} \right)^{1/3} \Omega_{gw} (\tau) \; . 
\end{equation}
According to the discussion above, the quantity on the r.h.s. of Eq.~\eqref{OmegaOmega} slowly varies with time reflecting the change of relativistic degrees of freedom in the primordial plasma. GWs with wavenumbers $k \lesssim k_{peak}$ receive the main contribution when they are on-horizon, i.e., $k/a \sim 2\pi H$, so that one should set $\tau \sim 2\pi/k$ in Eq.~\eqref{OmegaOmega}. On the other hand, the energy density of GWs with wavenumbers $k \gtrsim k_{peak}$ is mainly saturated at the times $\tau \sim \tau_{sc}$. In what follows, however, we will neglect this subtlety and set $\tau=\tau_{sc}$, which is justified if the duration of melting DW network operation is not very long, or frequency range of relevant experiments is not very broad.

\section{Optimising the system for lattice simulations}
\label{sec:optimisation}

To minimise numerical errors it is convenient to choose physical parameters and rescale variables in such a way that all the terms in equations to be solved on a lattice become of the order unity. In particular, we choose $\alpha=1$, 
so that the expectation value $\eta$ equals the Universe temperature, $\eta =T$. 
It is straightforward to properly rescale the results of our simulations for an arbitrary value of $\alpha$. We introduce dimensionless variables for the conformal time $\tau$, conformal momenta $k_l$, space conformal coordinates $x_l$, and scalar field $\chi$ as
\begin{equation}
\label{redefinitions}
 \sqrt{\lambda_{\chi}} \eta_i \tau \rightarrow \tau  \qquad 
\frac{k_l}{\sqrt{\lambda_{\chi}} \eta_i} \rightarrow k_l  \qquad   \sqrt{\lambda_{\chi}} \eta_i \, x_l \rightarrow x_l \qquad \frac{\chi}{\eta_i} \rightarrow \chi  \; .
\end{equation}
We choose to start simulations when the following equality is fulfilled:
\begin{equation}
\label{init}
H_i =\sqrt{\lambda_{\chi}}\, \eta_i =\sqrt{\lambda_{\chi}} \, T_i\; .
\end{equation}
We set to unity the initial dimensionless conformal time, $\tau_i=1$. 
In combinations with Eqs.~\eqref{redefinitions} and~\eqref{init}, this allows us to write for the scale factor $a(\tau)=\tau$. 

In dimensionless variables we can rewrite the equation of motion for the scalar field~\eqref{scalareq} as
\begin{equation}
\label{dimensionless}
\chi''+\frac{2}{\tau} \chi'-\frac{\partial^2 \chi}{\partial x^2_i}+\chi \cdot  (\chi^2 \cdot \tau^2 -1) =0 \; .
\end{equation}
It is important that the coupling constant $\lambda_{\chi}$ drops out of this equation. Consequently, as far as Eq.\,\eqref{dimensionless} is concerned, we can set $\lambda_{\chi}$ to an arbitrary value. We choose $\lambda_{\chi} =0.03$. As a result, the condition~\eqref{init} fixes 
the initial temperature to be $T_i \approx 1.3 \cdot 10^{17}~\mbox{GeV}$,
while the Hubble rate is given by $H_i \approx 2.3 \cdot 10^{16}~\mbox{GeV}$. We have used the expression of the Hubble rate during radiation domination
$H(T)=\sqrt{\pi^2 g_* (T)/90} \cdot T^2/M_{Pl}$, and we have set $g_* (T)=100$. Such large values of $T_i$ and $H_i$ are not realistic in the post-inflationary Universe, 
but this is not important: upon a proper rescaling we will be able to obtain results for an arbitrary $T_i$. 

One can further simplify Eq.~\eqref{dimensionless} by switching to the field variable $s=a\chi$. In terms of the field $s$ and dimensionless variables, the equation of motion takes the form 
\begin{equation}
s''-\partial^2_i s +s (s^2-1) =0 \; ,
\end{equation}
which is just a dimensionless version of Eq.~\eqref{Minkowski}. This equation describes dynamics of the scalar field with a unit vacuum expectation value in Minkowski spacetime. Nevertheless, when performing simulations, we choose to deal with the original field $\chi$ evolving in the radiation-dominated background. 

In the case of melting DWs, both the width $\delta_w (\tau)=\sqrt{\frac{2}{\lambda_{\chi}}} \frac{1}{\eta (\tau)}$ and the lattice spacing grow linearly with the scale factor $\propto a$. Therefore, if initially the width $\delta_{w,i}$ is larger than the initial lattice spacing $L_i/N$, the same will hold true at all the times while the DW network exists. We impose the condition 
\begin{equation}
\label{onelimit}
\delta_{w,i} =\sqrt{\frac{2}{\lambda_{\chi}}}  \frac{1}{\eta_i} =\frac{\kappa L_i}{N} \;,
\end{equation}
where we have introduced the parameter $\kappa$, which controls how much larger the wall width should be compared to the lattice spacing. 
In what follows, we adopt $\kappa=3.5$, unless otherwise stated.

Applying Eq.\,\eqref{init} one can rewrite Eq.~\eqref{onelimit} as 
\begin{equation}
\label{onelimitmod}
H (\tau_i) =\sqrt{\lambda_{\chi}} \eta_i=\frac{\sqrt{2} N}{\kappa L} \; .
\end{equation}
With $\lambda_{\chi}$ and $T_i$ fixed, this equation defines the constant comoving box size 
$L=L_i$ (recall that $a(\tau_i)=1$) given a particular grid number $N$.
We should also control that the simulation box always remains larger than the horizon volume. We terminate the simulations, when this condition is getting violated. 
In practice, we define the final time of simulations $\tau_f$ such that the box size $L$ is twice larger than the horizon size:
\begin{equation}
\label{twolimit}
H^{-1} (\tau_f) =\frac{a(\tau_f) \cdot L}{2} \; .
\end{equation}
From Eqs.~\eqref{onelimitmod} and~\eqref{twolimit}, we get
\begin{equation}
\label{time_covered}
\tau_f =\frac{N}{\sqrt{2} \kappa} \; .
\end{equation}
This proves the point made in the introduction that the time available for simulations $\Delta\tau$ grows linearly with the grid number $N$. 
This is contrary to the situation with constant tension DWs, where the corresponding time grows as $\sqrt{N}$.

\section{Numerical analysis of melting domain wall evolution}
\label{sec:dwevolution}

In this section, we study numerically evolution of the melting DW network using the code \CLns ~and following the lines of our previous study of constant tension 
DW network\,\cite{Dankovsky:2024zvs}. Compared to the case of constant tension DWs, evolution of melting walls is quite sensitive to the amplitude of initial scalar fluctuations. As in Ref.~\cite{Dankovsky:2024zvs} we start evolution of the scalar field $\chi$ from vacuum and thermal initial conditions. When setting initial conditions, we loosely assume that the field $\chi$ is massless and lives in the flat spacetime. This is irrelevant, as soon as we are not interested in particular initial conditions, but rather how the system evolves from {\it different} starting points. In the case of vacuum initial conditions, to make the initial scalar field variance finite, we impose the cutoff at some momentum $k_{cut}$, so that 
\begin{equation}
\label{invacuum}
\langle \delta \chi^2 \rangle_{vac} =\int^{k_{cut}}_0 \frac{dk}{k} \cdot \left(\frac{k}{2\pi} \right)^2 =\frac{k^2_{cut}}{8\pi^2} \; .
\end{equation}
We will also investigate vacuum initial conditions with the lattice UV cutoff, which means in practice that one sets $k_{cut}=k_{max}$, where $k_{max}=\sqrt{3}\pi N/L$ is the maximum momentum resolved on the lattice. 
In the case of thermal initial conditions, the initial scalar field variance reads 
\begin{equation}
\label{inthermal}
\langle \delta \chi^2 \rangle_{thermal} =\frac{1}{2\pi^2}\int^{\infty}_0 \frac{dk k}{e^{\frac{k}{T_i}}-1} =\frac{T^2_i}{12} \; .  
\end{equation}
In what follows, we also apply the cutoff to thermal initial conditions, so that the integration in Eq.~\eqref{inthermal}  extends to some finite $k_{cut}$ rather than to infinity (or $k_{max}$). In particular, if $k_{cut} \ll T_i$, one can write 
\begin{equation}
\label{inthermal2}
\langle \delta \chi^2 \rangle_{thermal} =\frac{k_{cut} T_i}{2\pi^2} \; .
\end{equation}
When performing numerical simulations, we consider a selection of values of $k_{cut}$, which enable us to investigate thoroughly dependence of the system on initial conditions. 

\begin{figure}[!htb]
\begin{center}
    \includegraphics[width=0.4\textwidth]{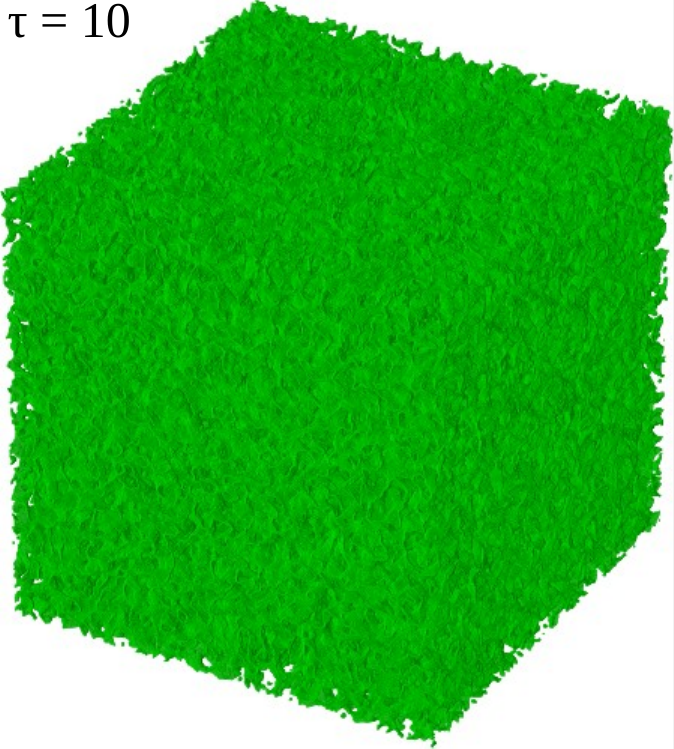} 
    \includegraphics[width=0.4\textwidth]{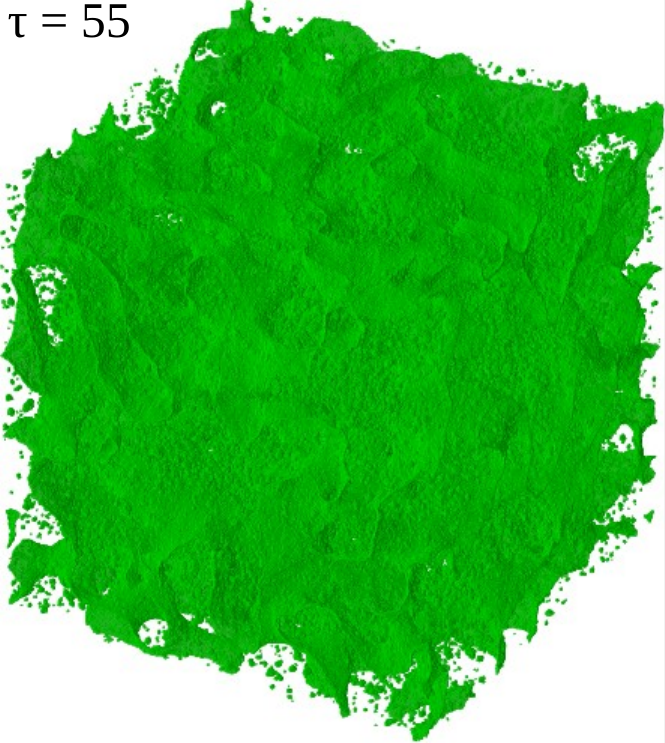} 
        \includegraphics[width=0.4\textwidth]{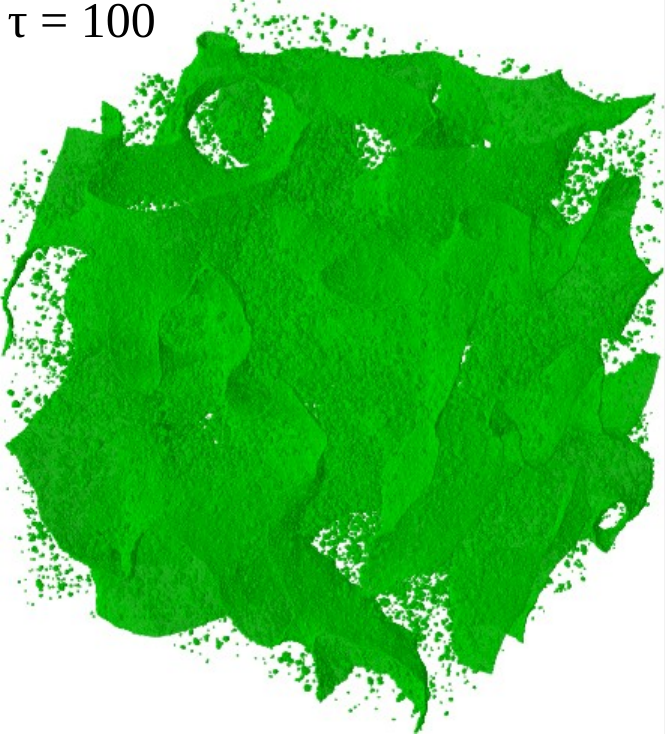} 
        \includegraphics[width=0.4\textwidth]{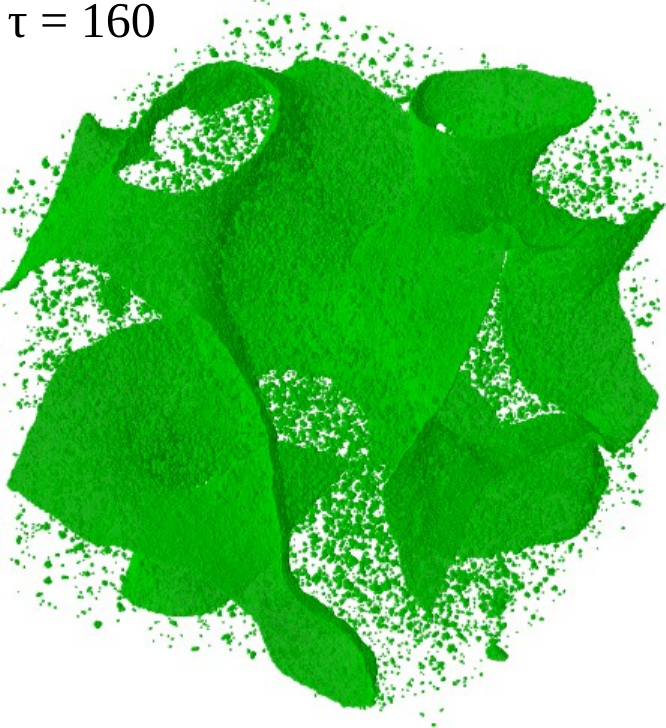} 
\end{center}
    \caption{Snapshots of melting DW evolution obtained with the $1024^3$ lattice in the case of vacuum initial conditions with $k_{cut}=1$. We use dimensionless units of Eq.~\eqref{redefinitions}.} \label{snapshots}
\end{figure}

Snapshots of melting DW network in the case of vacuum initial conditions with the cutoff $k_{cut}=1$ are shown in Fig.~\ref{snapshots}. As in the case of constant tension DWs (see Fig.~2 in Ref.~\cite{Dankovsky:2024zvs}), we observe a single long wall stretching throughout the simulation box surrounded by small closed walls at sufficiently late times. The long wall becomes smoother with time, which is a manifestation of the scaling law (see below for more details). However, compared to the case of constant tension DWs, the number density of closed melting walls appears to be considerably larger. We quantify this statement in a short while.

\begin{figure}[!t]
    \includegraphics[width=\textwidth]{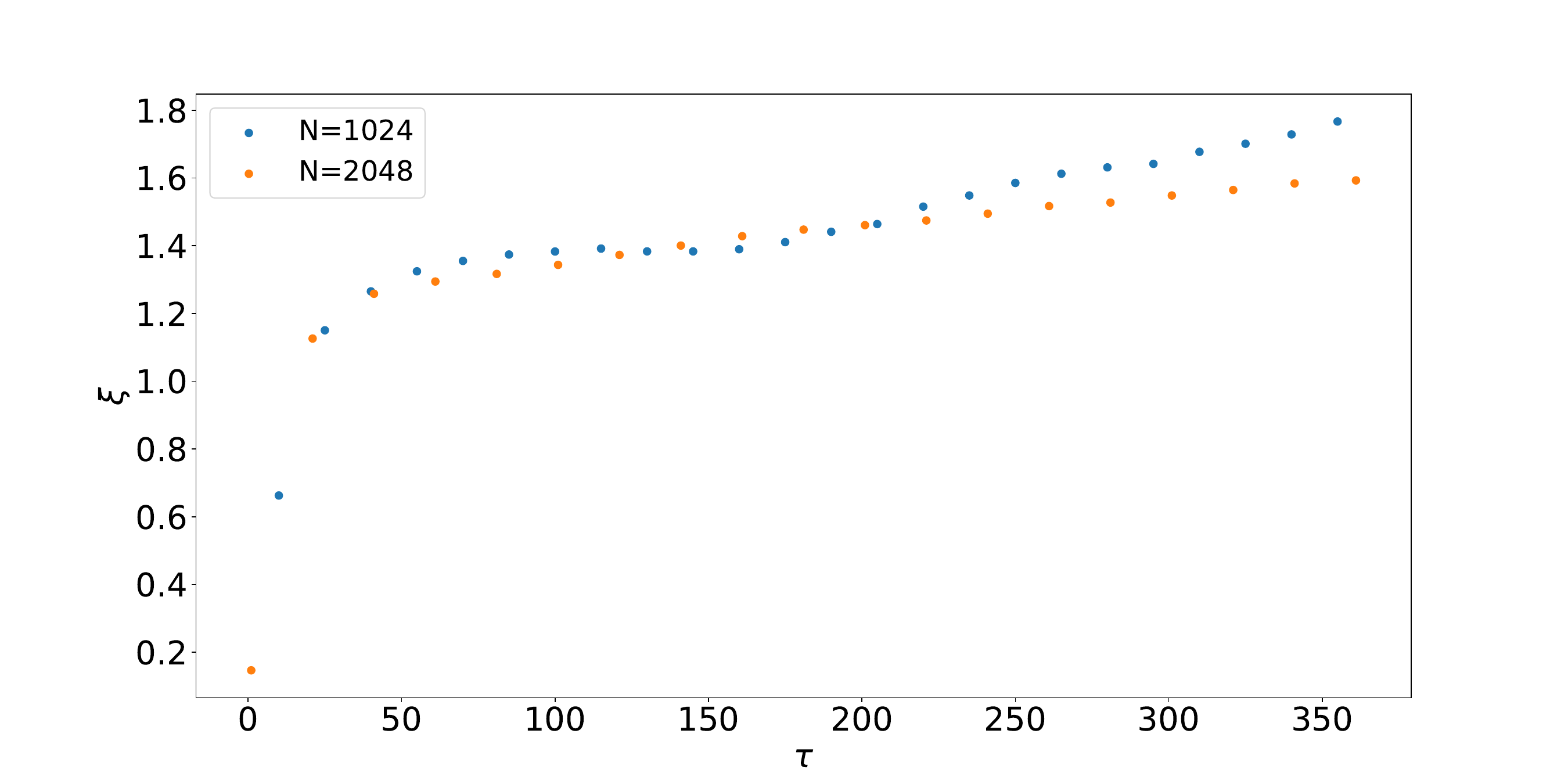} 
    \caption{Scaling parameter $\xi$ obtained with the $1024^3$ and $2048^3$ lattices in the case of vacuum initial conditions with the cutoff $k_{cut}=1$. One full simulation for each lattice is analysed. We use dimensionless units of Eq.~\eqref{redefinitions}.} \label{scaling_vacuum}
\end{figure}

One of the main quantitative characteristics of DWs is their area. The latter is recovered from lattice simulations using the estimator~\cite{Press:1989yh}: 
\begin{equation}
\label{links}
S=\Delta x^2 \sum_{links} \delta \frac{|\nabla \chi |}{|\chi_{,x}|+ |\chi_{,y}|+|\chi_{, z}|} \; ,
\end{equation}
where the links correspond to the pairs of grid points, and $\Delta x$ is the comoving distance between the grid points. In the sum we keep only contributions from the links such that the sign of the field $\chi$ changes at the grid points. This is regulated by the quantity $\delta$, which takes the values $\delta =1$ or $\delta =0$, if the scalar $\chi$ changes its sign or not, respectively. Evolution of DW area is shown in Fig.~\ref{scaling_vacuum} in terms of dimensionless area parameter $\xi$ defined in Eq.~\eqref{scalingdef0} for the case of relatively small scalar fluctuations modelled by means of vacuum initial conditions with the cutoff $k_{cut}=1$. We observe that the network reaches the scaling regime, 
i.e., the parameter $\xi$ takes on a constant value:
\begin{equation}
\xi_{sc} \approx 1.4 \; .
\end{equation}
The scaling starts roughly at the time
\begin{equation}
\label{scalingstart}
\tau_{sc} \simeq 25 \tau_i \; .
\end{equation}
This is considerably later compared to the case of constant tension walls, where $\tau_{sc} \simeq 5\tau_i$~\cite{Dankovsky:2024zvs}. Physically it can be more relevant to formulate the moment of settling to scaling in terms of the ratio $\delta_{w}/H^{-1}$. Notably, for both types of DWs the scaling starts when 
the wall width $\delta_{w}$ becomes about $5 \% $ of the Hubble radius $H^{-1}$. Since the width of melting DWs grows with time, though slower than the cosmological horizon, 
it takes longer for the system to reach the scaling solution in that case. One can see from Fig.~\ref{scaling_vacuum} that the scaling parameter $\xi$ obtained with $1024^3$ and $2048^3$ lattices starts to grow at late simulation times, which manifests a slight breaking of the scaling law. However, we attribute this growth, which is more prominent in the case of $1024^3$ lattice, to a finite size of the simulation box becoming comparable to the size of simulated Hubble patches by the end of simulations. Indeed, the analogous growth starts even earlier for the $512^3$ lattice (see the top left panel of Fig.~\ref{scaling_no}) assuming the same initial conditions and model parameters, because the simulation box must be chosen smaller in that case.

\begin{figure}[!htb]
    \includegraphics[width=0.5\textwidth]{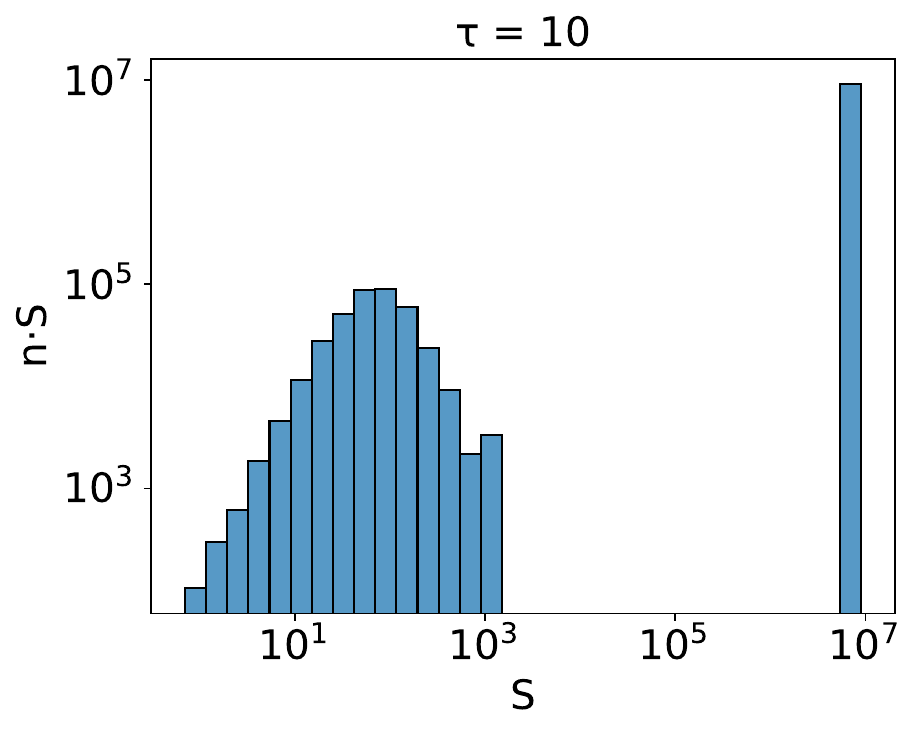} 
        \includegraphics[width=0.5\textwidth]{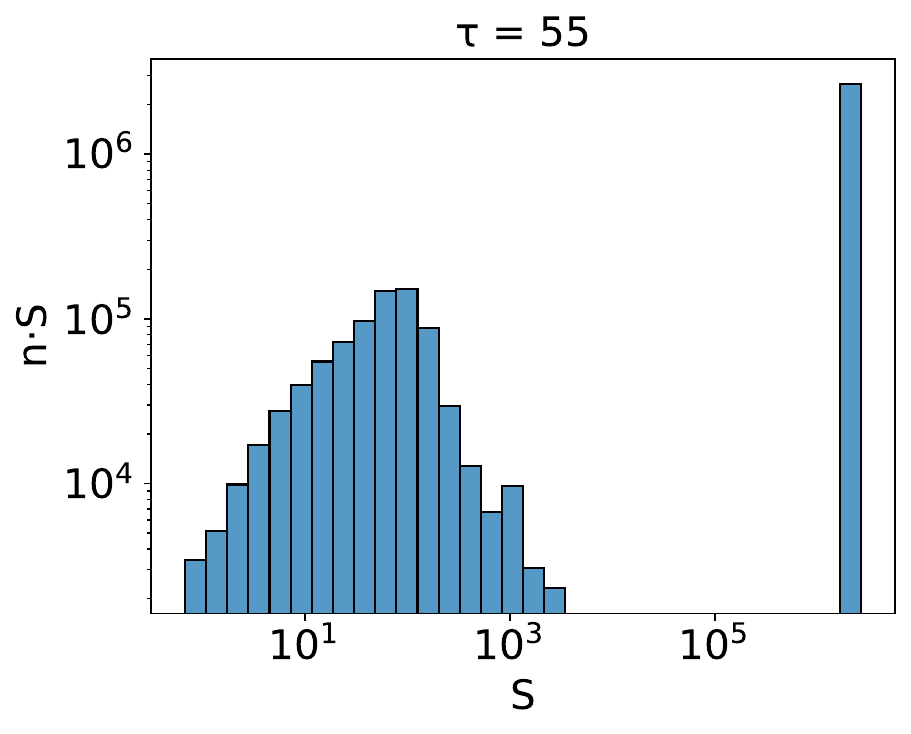} 
            \includegraphics[width=0.5\textwidth]{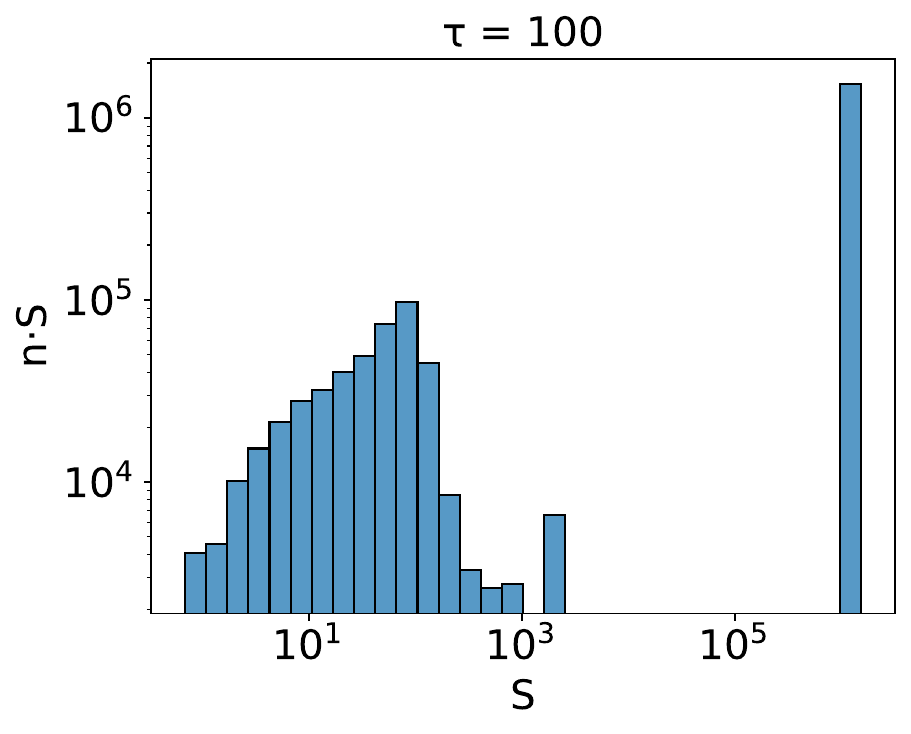} 
                \includegraphics[width=0.5\textwidth]{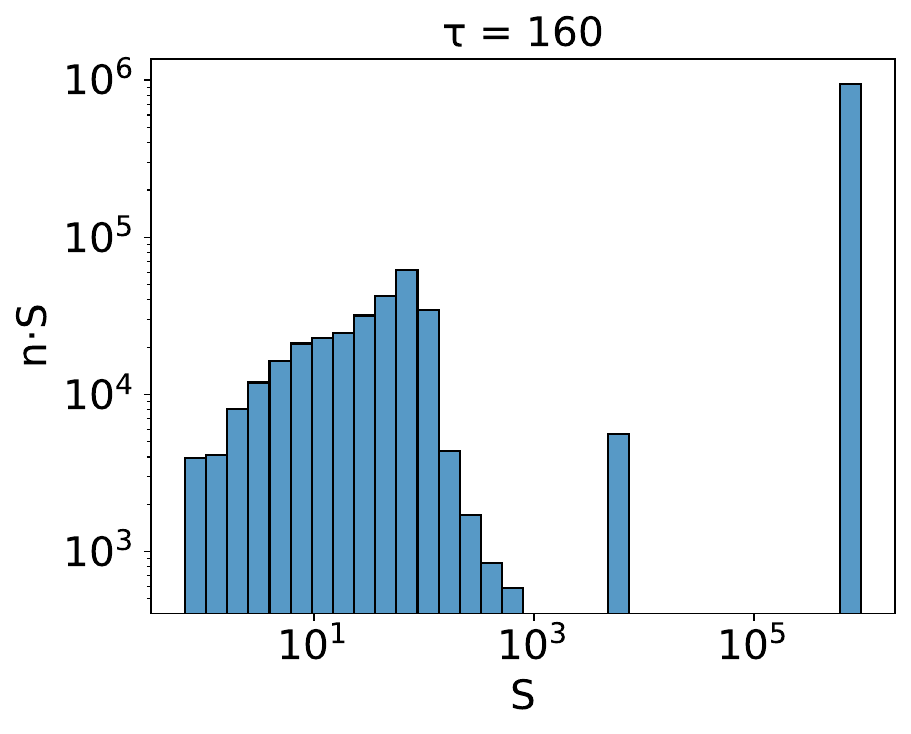} 
    \caption{Histograms showing distribution of melting DWs by the area $S$ in the case of vacuum initial conditions with the cutoff $k_{cut}=1$. The height of each column corresponds to the total area of all DWs in the small range $(S, S+\Delta S)$, where $\Delta S$ is the width of a column. Only one simulation with the $1024^3$ lattice has been performed for this plots. Dimensionless units of Eq.~\eqref{redefinitions} are used.} \label{Histograms}
\end{figure}

The actual difference between the two types of DWs is the distribution of the total area over a long wall and smaller closed walls. As it is clear from Fig.~\ref{Histograms}, 
\begin{figure}[!htb]
    \includegraphics[width=0.5\textwidth]{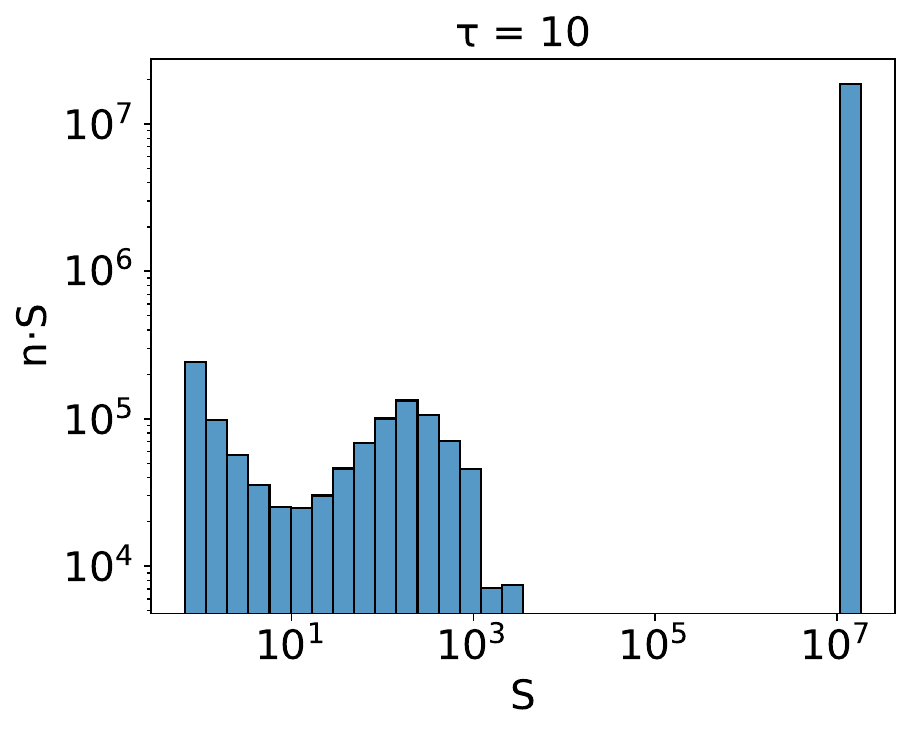} 
        \includegraphics[width=0.5\textwidth]{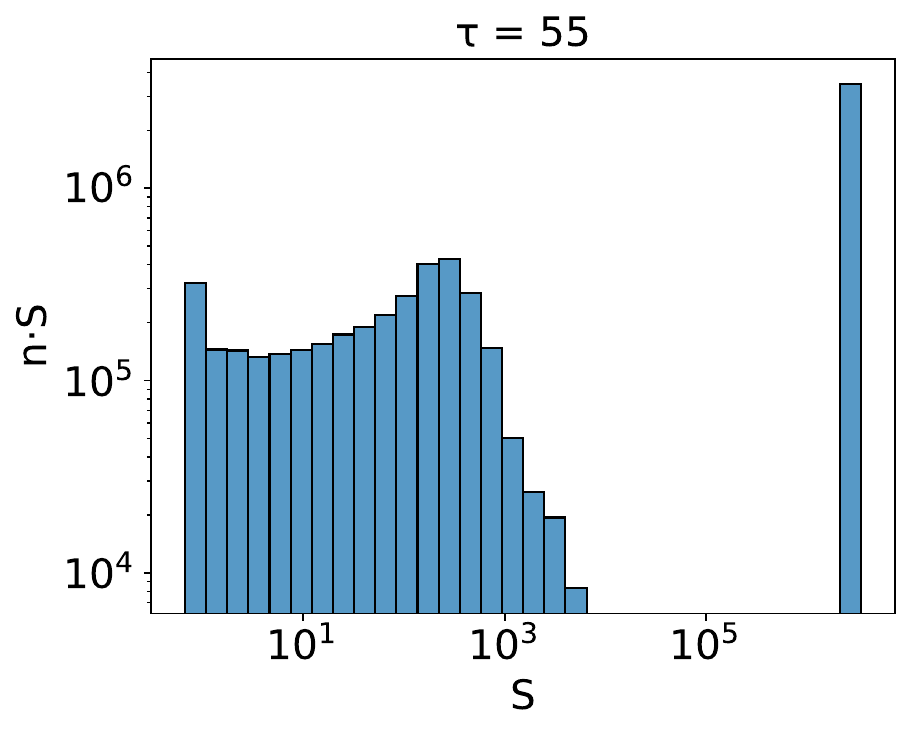} 
            \includegraphics[width=0.5\textwidth]{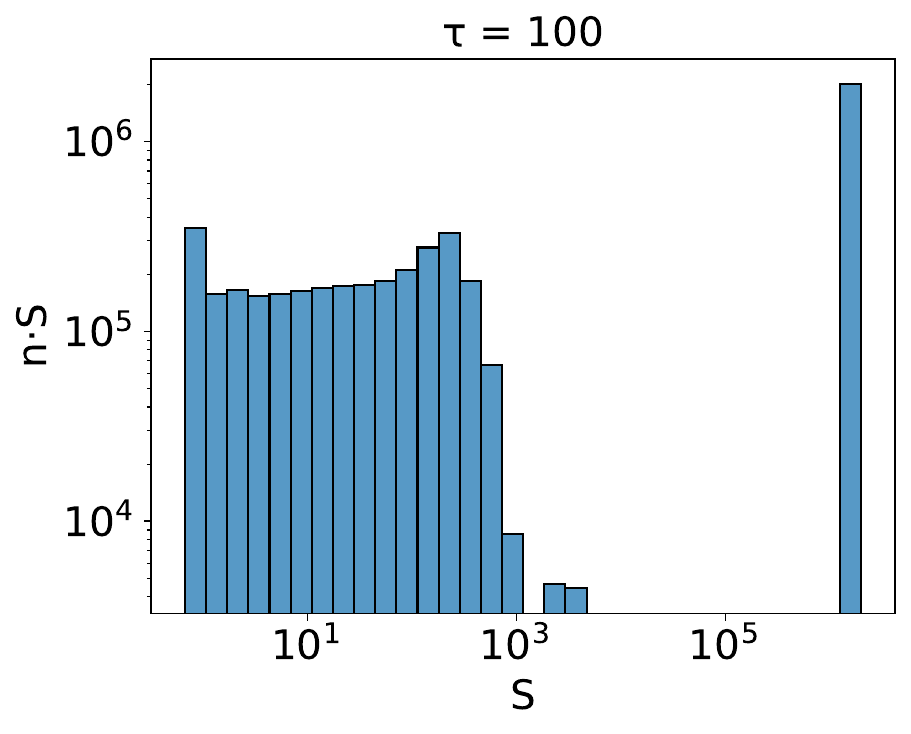} 
                \includegraphics[width=0.5\textwidth]{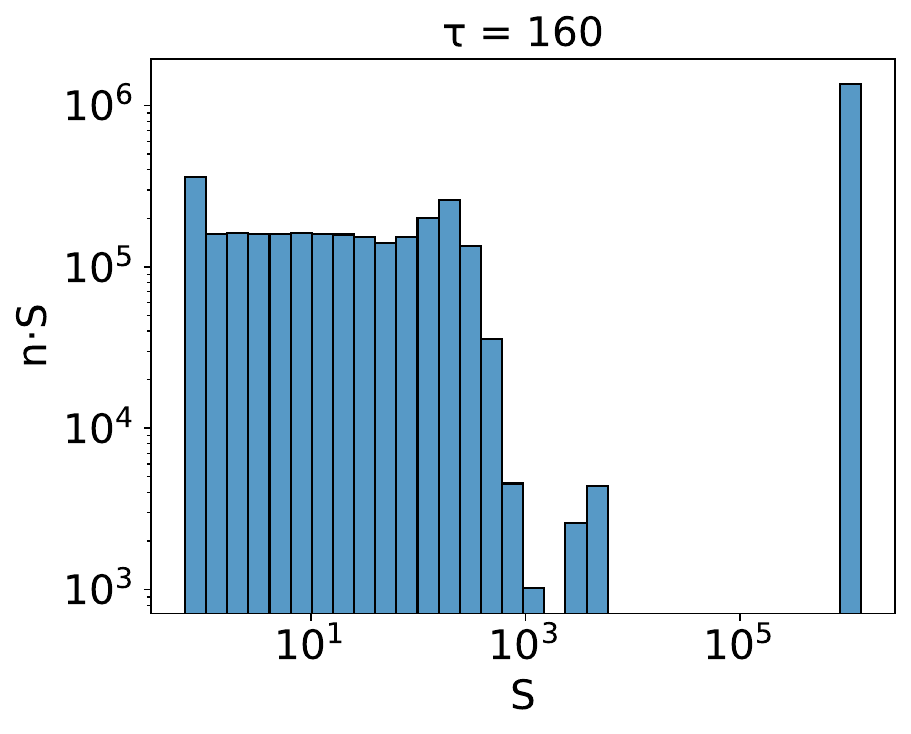} 
    \caption{The same as in Fig.~\ref{Histograms}, but assuming thermal initial conditions (with the lattice UV cutoff).} \label{Histograms_thermal}
\end{figure}
dynamics of the network is dominated by a single long wall stretching throughout the simulation box. 
Nevertheless, the contribution of closed walls is non-negligible, --- in a good agreement with the snapshots in Fig.~\ref{snapshots}. Namely, one finds for the corresponding ratio
\begin{equation}
\label{r}
r \equiv \frac{S (closed~ walls)}{S (long~wall)} \simeq 0.3 \; ,
\end{equation}
which has been cross-checked between $512^3$ and $1024^3$ lattices.
 This suggests that closed walls may play an important, if not crucial role, in reaching the scaling regime. 
On the other hand, the analogous ratio $r$ is an order of magnitude smaller compared to Eq.~\eqref{r} in the case of constant tension DWs, as it follows from the analysis of Ref.~\cite{Dankovsky:2024zvs}.  
For a viable explanation of such a difference, it was suggested in Ref.~\cite{Vilenkin} that the Universe expansion facilitates wall smoothing. 
At the same time recall that melting DWs effectively do not experience a cosmic expansion: their evolution at radiation domination is equivalent to that of constant tension walls 
in Minkowski spacetime. Thus, it is conceivable that the only way for melting DWs to reach scaling is through an extensive formation of closed walls eventually being dissolved into particles.

There is another crucial difference between the two types of DWs. Namely, unlike in the case of constant tension walls, the appearance of scaling regime in the case of melting walls depends on the amplitude of the initial scalar fluctuations. This is evident from Fig.~\ref{scaling_no}, 
\begin{figure}
    \includegraphics[width=0.98\textwidth]{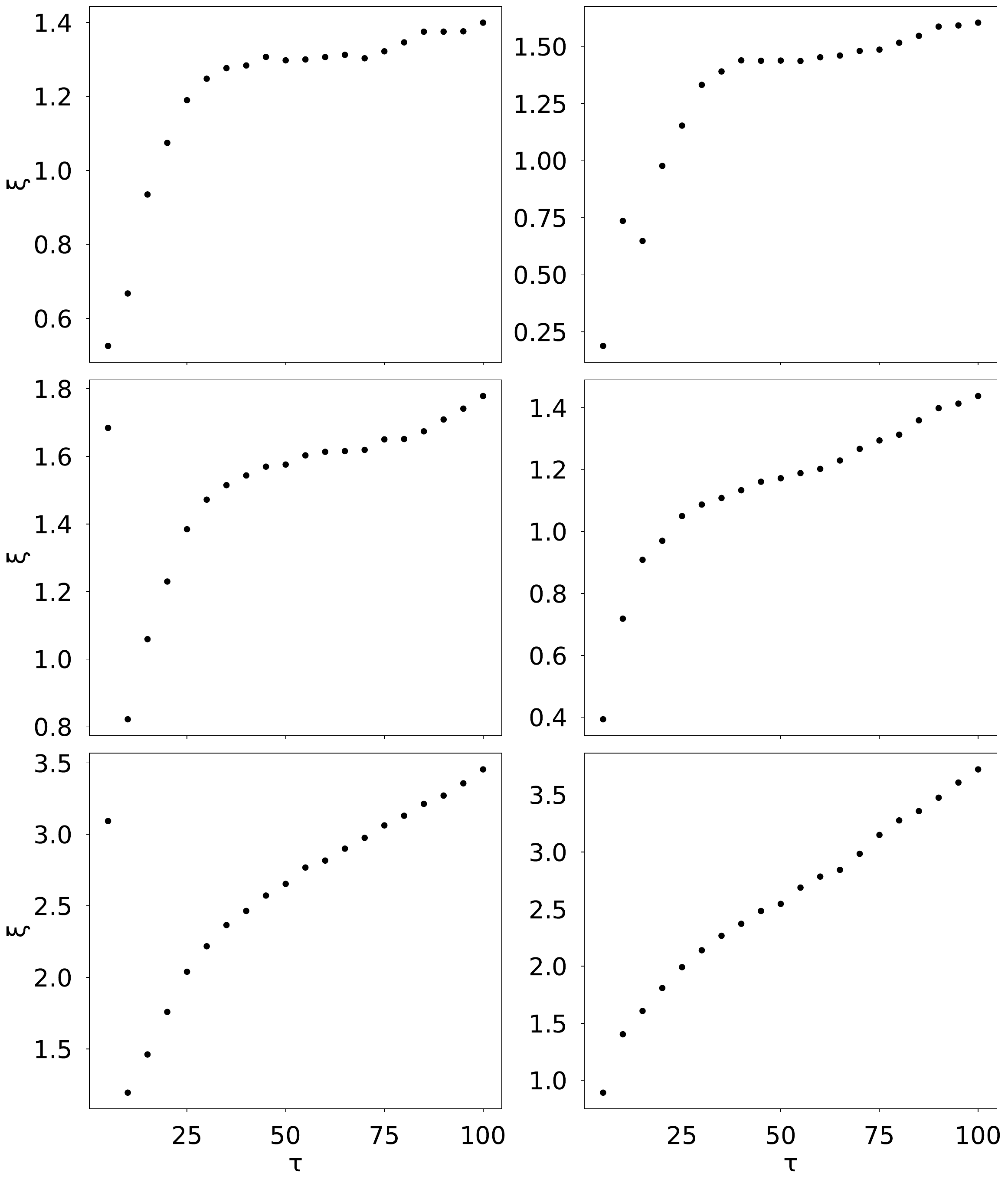}
    \caption{Left panel. Scaling parameter $\xi$ obtained with $512^3$ lattice in the case of vacuum initial conditions with the cutoff $k_{cut}=1$ (top), $k_{cut}=6$ (middle), and with the lattice UV cutoff (bottom). Right panel. Same but assuming thermal initial conditions with the cutoff $k_{cut}=0.3$ (top), $k_{cut}=1$ (middle), and with the lattice UV cutoff (bottom). One simulation has been performed for each plot. We adopt dimensionless units defined in Eq.~\eqref{redefinitions}.}  \label{scaling_no}
\end{figure}
where we again choose vacuum initial conditions, but we increase the scalar amplitude by increasing the cutoff $k_{cut}$. The similar picture occurs in the case of thermal initial conditions with the scaling parameter demonstrated in Fig.~\ref{scaling_no}. While for thermal initial conditions the scaling law is not obeyed (see the right plot in Fig.~\ref{scaling_no}), imposing the artificial cutoff on the initial black body distribution of the scalar field, one regains the scaling behaviour. We have observed that the melting DW network starts violating the scaling roughly at the cutoff $k_{cut} \sim 5$ and $k_{cut} \sim 1$ in the case of vacuum and thermal initial conditions, respectively, see Fig.~\ref{scaling_no}. Using Eqs.~\eqref{invacuum} and~\eqref{inthermal2}, these can be converted into the borderline amplitude of the scalar fluctuation, at which the transition between the scaling and non-scaling behaviour happens (in units of $\eta_i$):
\begin{equation}
\delta \chi_{cr} \sim 0.1 \; .
\end{equation}
Namely, for $\sqrt{\langle \delta \chi^2 \rangle} \ll \delta \chi_{cr}$ initially, one observes the scaling behaviour, while for $\sqrt{\langle \delta \chi^2 \rangle} \gg \delta \chi_{cr}$, the scaling law is grossly violated.

\begin{figure}[!htb]
    \includegraphics[width=\textwidth]{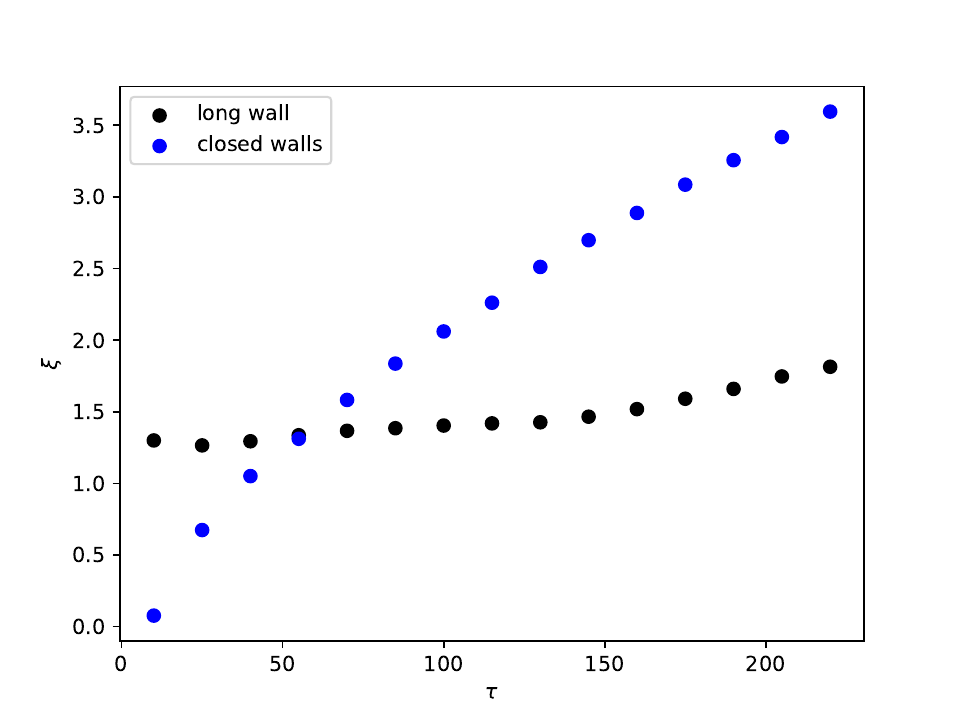} 
    \caption{Evolution of the scaling parameter $\xi$ is shown separately for a long wall (black dots) and closed walls (blue dots) in the case of thermal initial conditions (with the lattice UV cutoff). Simulations are performed on $1024^3$ lattice. Dimensionless units of Eq.~\eqref{redefinitions} have been used.} \label{semi_scaling}
\end{figure}

Interestingly, in the situation with a large initial scalar fluctuation gross violation of the scaling behaviour observed is due to an abundant production of small closed walls. At the same time, the long wall area still approximately obeys the scaling law, see Fig.~\ref{semi_scaling}. Note, however, that the minimal comoving area of closed walls cannot get smaller than $\sim \pi \delta^2_{wall}$. Yet we have checked with histograms that the area of closed DWs breaks this limit in the situation with no scaling. This suggests a potential explanation for the breaking of the scaling law: it is possible that the estimator~\eqref{links} simply miscounts large fluctuations of the scalar field at small scales as closed DWs. 

Besides this explanation for a qualitative change seen in Fig.~\ref{scaling_no}, which is of systematic origin, we would like to mention another one of the physical origin. Namely, the scalar field $\chi$ gets an opportunity to cross the potential barrier, if its initial fluctuations are large. 
Such a tendency is not prominent in the case of constant tension walls, 
because in that case large oscillations of the scalar field triggered by its large initial amplitude, decay with time due to Hubble friction. The similar effect is present in the case of melting DWs, but it is fully compensated by the time decreasing expectation value, as it follows from the equivalence of melting DWs at radiation domination and constant tension DWs in the flat spacetime (where both the height of the potential barrier and the scalar field fluctuation amplitudes remain constant in time). Consequently, the Hubble damping does not prevent the scalar field from crossing the barrier in the case of melting DWs. Nevertheless, it is unclear how the barrier crossing leads to an accumulation of small DWs observed in Fig.~\ref{scaling_no} rather than their disruption.

\section{Numerical analysis of gravitational wave emission}
\label{sec:gwnumerics}

\begin{figure}[!htb]
    \includegraphics[width=\textwidth]{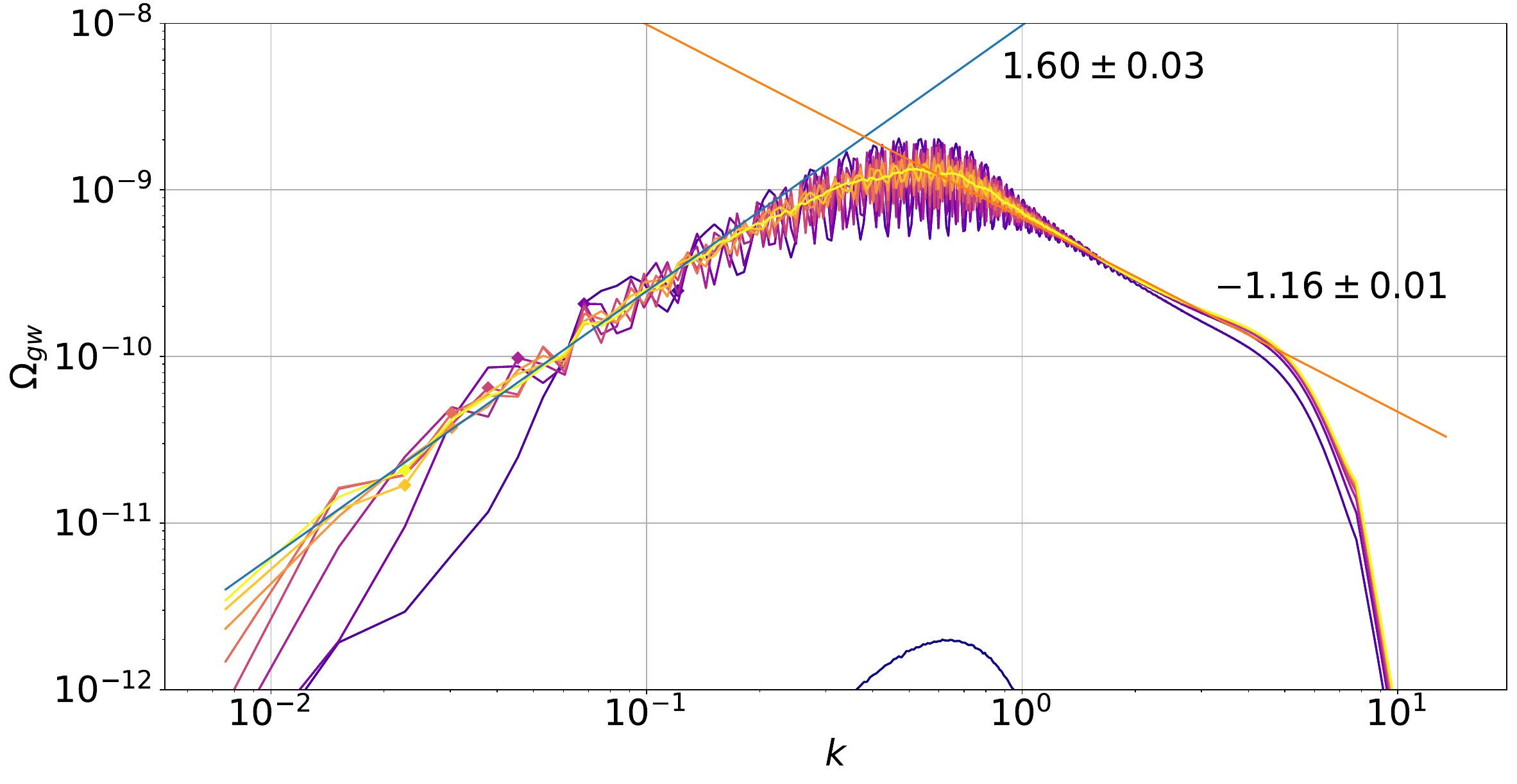} 
    \caption{Spectrum of GWs produced by the network of melting DWs in the case of vacuum initial conditions with the cutoff $k_{cut}=1$ in dimensionless units defined in Eq.~\eqref{redefinitions}. One simulation is performed on $2048^3$ lattice. Brighter colors correspond to the spectrum at later times. The initial and final conformal times of the simulation are set at $\tau_i=1$ and $\tau = 405$ (shown with the yellow line), respectively. Straight lines demonstrate slopes of GW spectrum in its close-to-maximum infrared and ultraviolet parts at $\tau=405$; corresponding spectral indices are also shown. We have set $\alpha=1$, $\lambda_{\chi}=0.03$, $g_* (T_{sc})=100$, and $T_i \approx 1.3 \cdot 10^{17}~\mbox{GeV}$, when performing simulations. For arbitrary $\alpha$, $\lambda_{\chi}$, $g_* (T_{sc})$, and $T_i$, one should multiply values on the plot by $\alpha^6 \cdot (\lambda_{\chi}/0.03) \cdot (T_i/1.3 \cdot 10^{17}~\mbox{GeV})^2 \cdot (100/g_*(T_{sc}))$, see Eq.~\eqref{peaknumerics}.} 
    \label{spectrum_vacuum}
\end{figure}

\begin{figure}[!htb]
    \includegraphics[width=\textwidth]{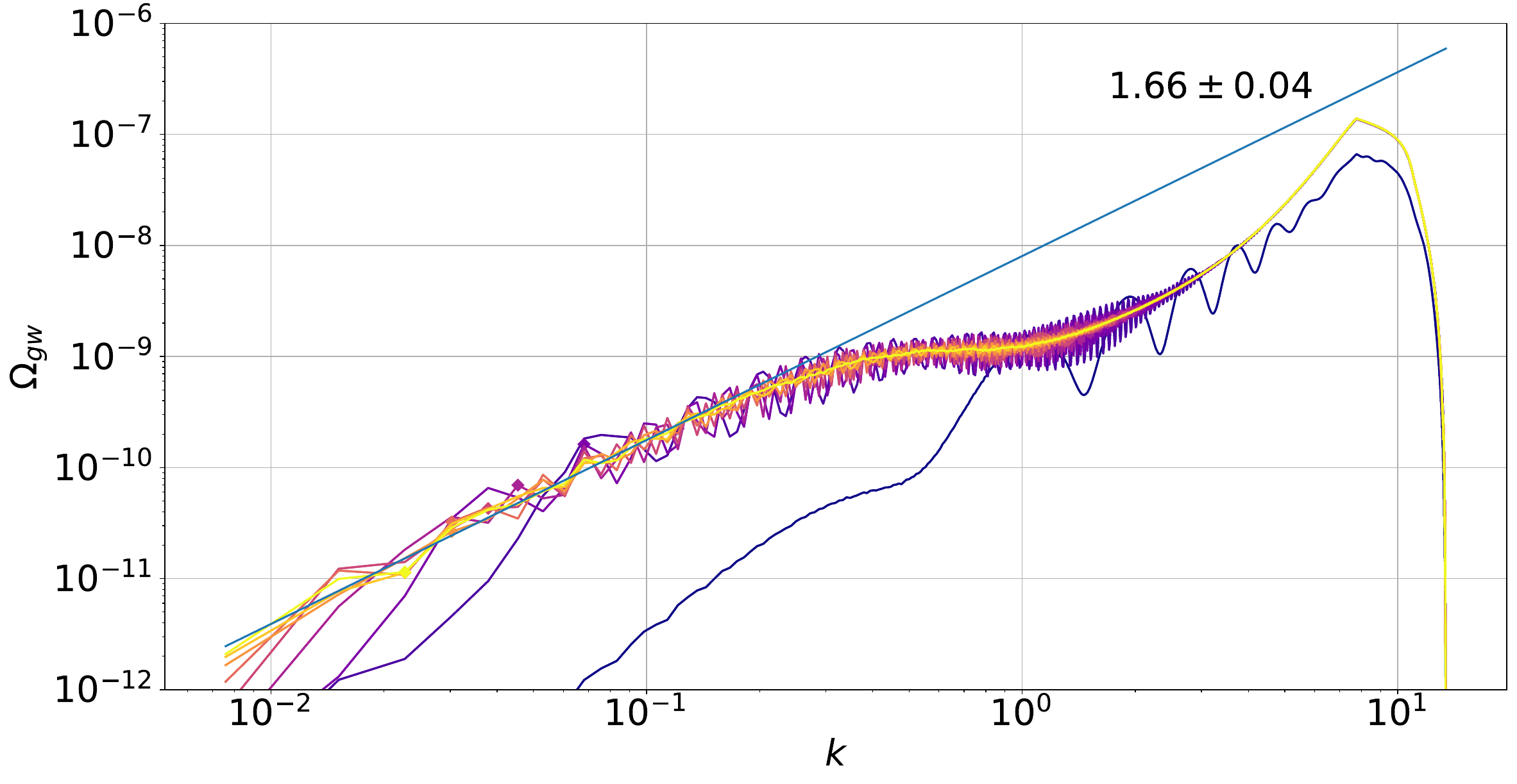} 
    \caption{The same as in Fig.~\ref{spectrum_vacuum}, but in the case of thermal initial conditions (with the lattice UV cutoff).} \label{spectrum_thermal}
\end{figure}

In numerical simulations, we obtained two types of GW spectra shown in Fig.~\ref{spectrum_vacuum} and Fig.~\ref{spectrum_thermal} depending on the initial conditions for the scalar field. This is in accordance with the discussion in the previous section, where we demonstrated that evolution of melting DWs may or may not reach the scaling regime depending on the initial scalar fluctuations.

We start our analysis with Fig.~\ref{spectrum_vacuum}
showing the spectrum of GWs for vacuum initial conditions with a cutoff $k_{cut}=1$, in which case melting DWs obey the scaling law. The spectrum of GWs peaks at the wavenumber $k_{peak} \simeq 4\pi/\tau_{sc}$ corresponding to frequency $F_{peak}\simeq 2H (\tau_{sc})$ at the onset of scaling, when most energetic GWs are emitted. The present day peak frequency reads
\begin{equation}
\label{frtemp}
f_{peak} \simeq 2H (\tau_{sc}) \cdot \frac{a_{sc}}{a_0} \approx 15~\mbox{nHz} \cdot  g^{1/6}_* (T_{sc}) \cdot \left(\frac{T_{sc}}{100~\mbox{MeV}} \right)   \; .
\end{equation}
From Fig.~\ref{spectrum_vacuum}, one can also infer the value of the peak height, i.e., $\Omega_{gw, peak} \approx 10^{-9}$, which assumes 
a particular choice of model constants and initial conditions for the Universe, see the caption in Fig.~\ref{spectrum_vacuum}. One restores the value $\Omega_{gw, peak}$ for generic model constants from Eq.~\eqref{peakth}:
\begin{equation}
\label{peaknumerics}
\Omega_{gw, peak} \approx \frac{10 \lambda_{\chi} \alpha^6 \xi^2_{sc} G_N T^2_i}{g_* (T_{sc})} \cdot \left(\frac{\tau_i}{\tau_{sc}} \right)^2 \; .
\end{equation}
One can write more compactly $T_{sc} =T_i \tau_i/\tau_{sc}$ as in Eq.~\eqref{peakth}, but we would like to emphasise the role of the 
delayed onset of scaling resulting into the additional factor $(\tau_i/\tau_{sc})^2$ compared to the situation 
when the scaling is reached immediately upon the network formation. Note that GWs produced before the time $\tau_{sc}$ can be neglected. This is evident from the curve on the bottom of Fig.~\ref{spectrum_vacuum} corresponding to GWs generated by the time $\tau =5 \ll \tau_{sc}$.

Notably, the numerically obtained fractional energy GW energy density at peak~\eqref{peaknumerics} is only $\sim 2$ times larger than our theoretical 
estimate~\eqref{peakth}. It is also interesting to compare the results of numerical simulations shown in Fig.~\ref{spectrum_vacuum} with the theoretical estimate~\eqref{estimate} holding in the close-to-maximum infrared range $2\pi/\tau \ll k \ll 2\pi/\tau_{sc}$; we pick $k =0.07$. 
At this value Fig.~\ref{spectrum_vacuum} gives $\Omega_{gw} \approx 1.5 \cdot 10^{-10}$, while the estimate~\eqref{estimate} gives $\Omega_{gw} \sim 2.5 \cdot 10^{-10}$. Keeping in mind rough approximations when deriving the analytical formulas, we observe an excellent agreement 
between the latter and the numerical results.

On the left of the peak, the spectrum of GWs is described by 
\begin{equation}
\Omega_{gw} \propto k^{1.60 \pm 0.03}\, \qquad 2\pi/\tau \lesssim k \lesssim k_{peak} \; .
\end{equation}
This is in accordance with our discussion in Section~\ref{gw:theory} that a finite operation time of melting DWs 
(limited by formation time $\tau_i=1$ and the final time of simulations $\tau \approx 400$) leads to departures in the close-to-maximum infrared part from 
the spectral index $n=2$. On the right of the peak, in the close-to-maximum ultraviolet part of the spectrum, the spectral shape experiences a power-law decrease:
\begin{equation}
\label{rightspectrum}
\Omega_{gw} \propto k^{-1.16\pm 0.01} \qquad k_{peak} \lesssim k \lesssim 2\pi/\delta_{wall, c}\; ,
\end{equation}
where $\delta_{wall,c}$ is the comoving wall width. The spectral index in Eq.~\eqref{rightspectrum} is larger than in the analogous case of 
constant tension walls, i.e., $n \simeq -1.5$~\cite{Dankovsky:2024zvs}. Perhaps the difference can be attributed to the fact that the melting wall network 
exhibits more profound substructure in the form of small closed walls. In both cases of melting DWs and constant tension DWs, there is an exponential falloff due to a finite wall width, which is expected to start at $k_{wall} \simeq 2\pi /\delta_{wall,c}$, 
or $k_{wall} \simeq 4.5$ in dimensionless units, --- in a perfect agreement with Fig.~\ref{spectrum_vacuum}.

In the case of thermal initial conditions with parameters described in Section~\ref{sec:optimisation}, the spectrum of GWs is shown in Fig.~\ref{spectrum_thermal}, 
which is clearly distinct from Fig.~\ref{spectrum_vacuum} in the ultraviolet frequency range. In particular, the spectrum in Fig.~\ref{spectrum_thermal} grows beyond the momentum $k \simeq 4\pi/\tau_{sc}$, which previously corresponded to the peak, and now stands for the inflection point. Such a growth is not surprising, as it correlates with violation of the scaling law observed for large initial scalar fluctuations, as is the case of thermal conditions (assuming our choice of parameters).  
It is remarkable though that the spectral shapes are almost indistinguishable in the low frequency regime $k \lesssim 4\pi/\tau_{sc}$ on both Figs.~\ref{spectrum_vacuum} and~\ref{spectrum_thermal}. This can be explained by the fact that the infrared parts of the spectra mainly reflect dynamics of a long wall, which approximately obeys the scaling law in both cases. On the other hand, the high frequency part of the spectrum with $k \gtrsim 4\pi/\tau_{sc}$ is likely to be dominated by small closed walls/small scale fluctuations abundantly produced in the non-scaling case. 

\begin{figure}[!htb]
    \includegraphics[width=0.5\textwidth]{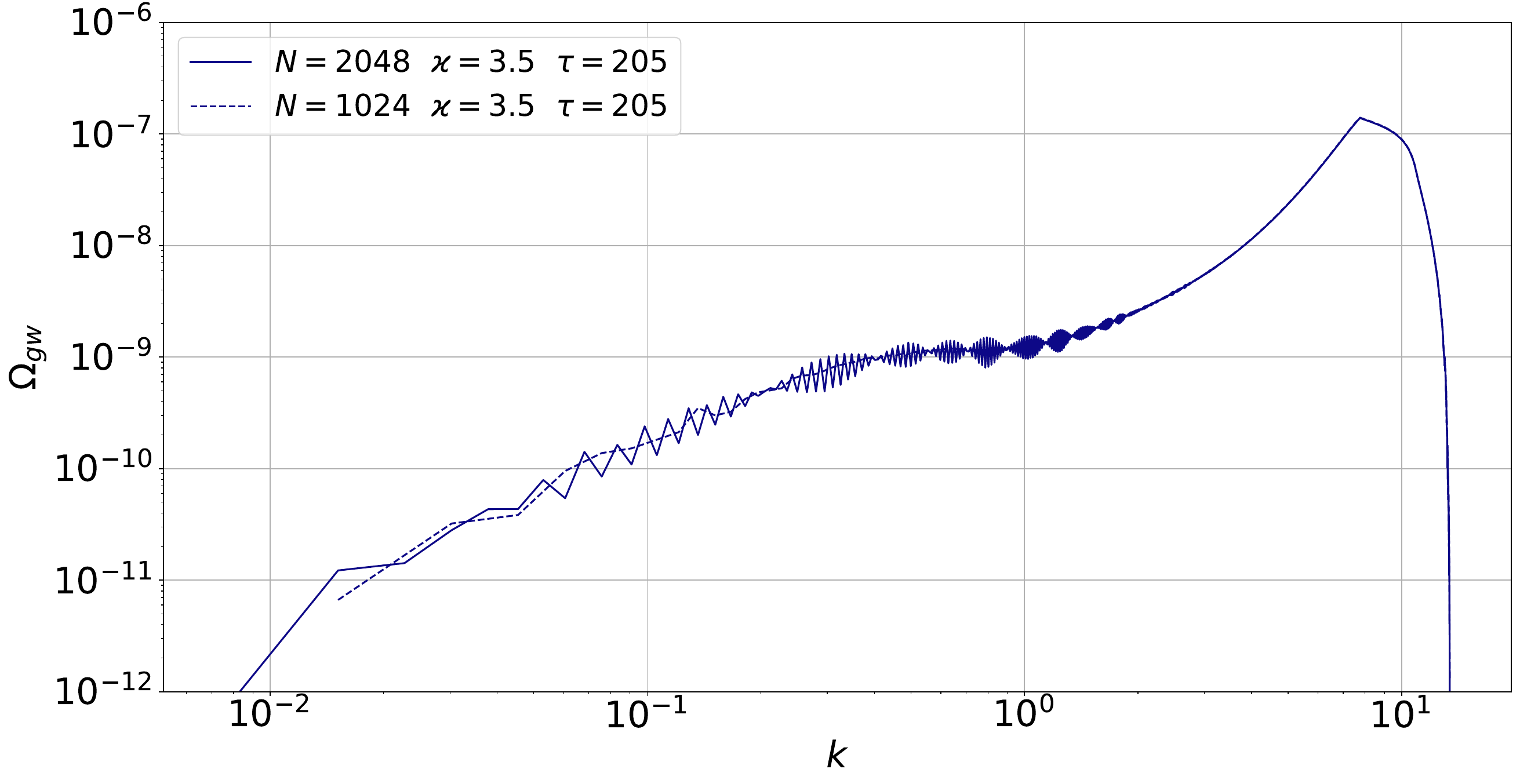} 
    \includegraphics[width=0.5\textwidth]{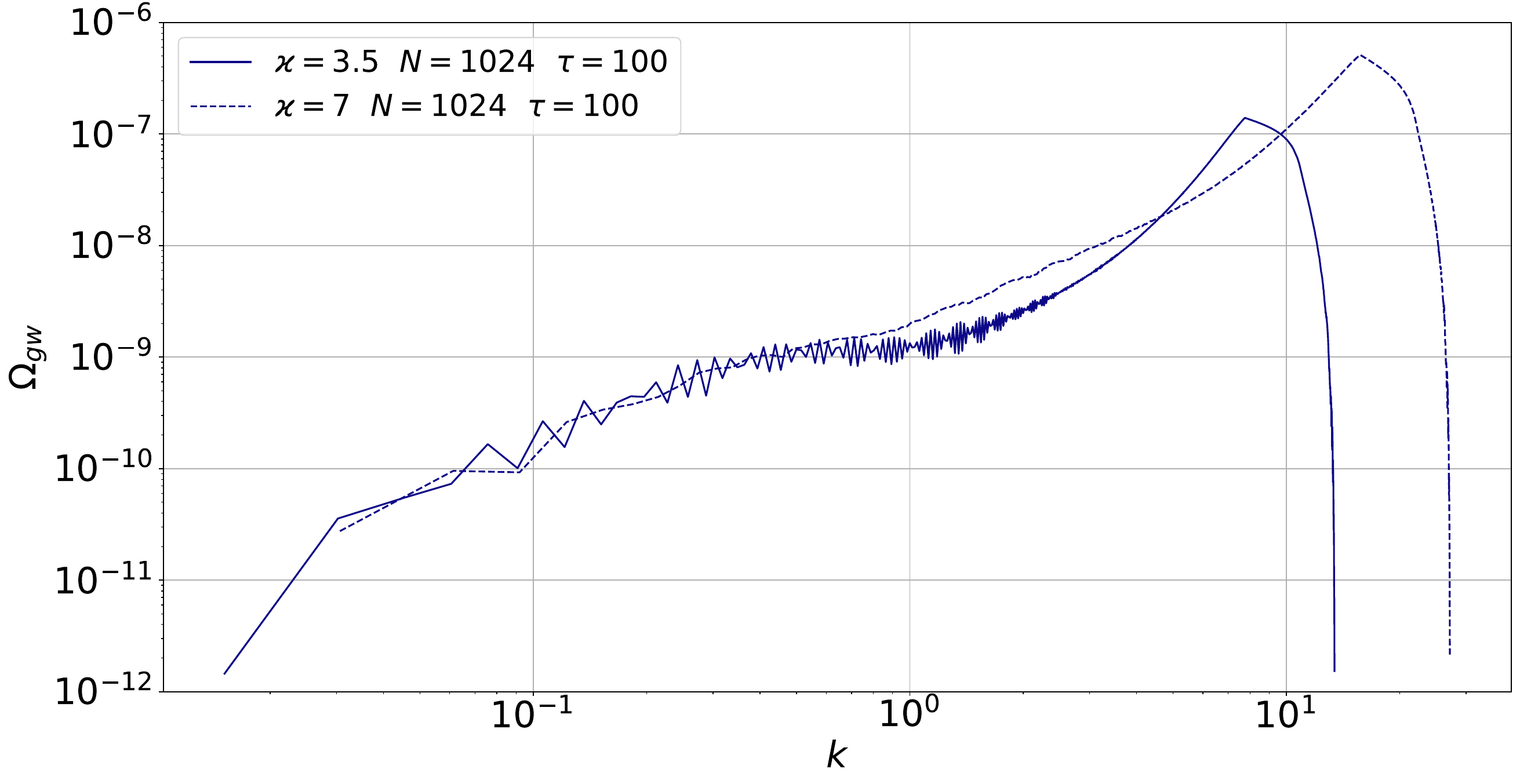} 
    \caption{Comparison of GW spectra obtained assuming thermal initial conditions with $2048^3$ and $1024^3$ lattices in the case of equal lattice spacing controlled by parameter $\kappa$ introduced in Eq.~\eqref{onelimit} (left) and different lattice spacings (right). Dimensionless units of Eq.~\eqref{redefinitions} have been used.} \label{comparison}
\end{figure}

It is questionable if the high frequency part of Fig.~\ref{spectrum_thermal} can be trusted. As it has been mentioned in the previous section, one can misinterpret closed DWs with scalar fluctuations of the non-topological origin at very small scales. Nevertheless, these scalar fluctuations themselves are physical, and so is the growth of the spectrum at $k \gtrsim 4\pi/\tau_{sc}$. This is confirmed by the fact that switching from $2048^3$ lattice to $1024^3$ lattice while keeping the lattice spacing intact 
does not impact our results, see the left plot of Fig.~\ref{comparison}. Nevertheless, the peak location in Fig.~\ref{spectrum_thermal} is most likely non-physical, as it is linked to the lattice spacing. Namely, changing the lattice spacing alters the peak location, as it is clear from the right plot of Fig.~\ref{comparison}. This is explained by the fact that scalar fluctuations are erased at scales below the lattice spacing, and hence they do not contribute to the GW spectrum.

To conclude, the low frequency part of the GW spectrum with $k \lesssim 4\pi/\tau_{sc}$ shows a remarkable robustness against initial conditions and thus represents a clear signature of melting DWs. 
The high frequency behaviour is trustworthy only in the case of initial conditions with a relatively small initial scalar fluctuation. 
Otherwise, the high frequency part becomes sensitive to the lattice spacing.

\section{Applications to particle physics}
\label{sec:applications}

In this section, we discuss a concrete particle physics scenario underlying the effective Lagrangian~\eqref{Lagrange}. 
Following Refs.~\cite{Babichev:2023pbf, Ramazanov:2021eya, Babichev:2021uvl}, we consider the model, which incorporates two coupled scalar fields: our singlet field 
$\chi$ constituting DWs, and the thermalised scalar multiplet $\phi$. The Lagrangian density describing the field $\chi$ is given by
\begin{equation}
\label{Lagrangemodel}
{\cal L}=\frac{(\partial \chi)^2}{2} -\frac{M^2_{\chi} \chi^2}{2}+\frac{g^2 \chi^2 \phi^\dagger \phi}{2} -\frac{\lambda_{\chi} \chi^4}{4} \; ,
\end{equation}
where $g^2$ and $\lambda_{\chi}$ are the coupling constant between the fields $\chi$ and $\phi$ and the quartic self-interaction 
of the scalar $\chi$, respectively. The scalar multiplet $\phi$ is generically described by ${\cal N}$ degrees of freedom, 
and schematically its Lagrange density reads
\begin{equation}
{\cal L}_{\phi}=\frac{1}{2}|D_{\mu} \phi|^2 -\frac{1}{2}m^2_{\phi} |\phi|^2-\frac{1}{4}\lambda_{\phi} |\phi|^4 +\mbox{interactions with other fields}\; .
\end{equation}
Here $m^2_{\phi}$ and $\lambda_{\phi}$ are the squared mass and the quartic self-coupling of the multiplet $\phi$. 
We do not write explicitly interactions of the field $\phi$ with other fields (except for $\chi$ in Eq.~\eqref{Lagrangemodel}), but they play a crucial role in keeping the particles $\phi$ in thermal equilibrium with hot plasma, 
at least at sufficiently early times. One can in principle identify  $\phi$ with the Higgs field, in which case  ${\cal N}=4$ and $m^2_{\phi}<0$. However, we find this option very restrictive from the viewpoint of GW emission, 
and therefore we keep ${\cal N}$ free. With no much loss of generality, we consider $m^2_{\phi}>0$ for simplicity. 
In the high temperature regime $T \gg m_{\phi}$, the field $\phi$ variance is given by
\begin{equation}
\label{variance}
\langle |\phi |^2 \rangle =\frac{{\cal N} T^2}{12} \; .
\end{equation}
For $g^2>0$, this variance feeds into the non-zero temperature-dependent expectation value $\eta$ of the field $\chi$ given by 
\begin{equation}
\eta =\pm \sqrt{\frac{{\cal N} g^2 T^2}{12\lambda_{\chi}} -\frac{M^2_{\chi}}{\lambda_{\chi}}} \; ,
\end{equation}
so that $Z_2$-symmetry is spontaneously broken at sufficiently high temperatures, and the DW network can be formed. With the assumption $g^2>0$, we run the risk of destabilising the two field system. 
The stability is protected by the quartic self-interaction terms, provided that the respective 
couplings obey the constraint: 
\begin{equation}
\lambda_{\chi} \lambda_{\phi} \geq g^4 \; .
\end{equation}
It is convenient to introduce the ``reduced self-interaction" coupling constant 
\begin{equation}
\label{bet}
\beta \equiv \frac{\lambda_{\chi}}{g^4} \geq \frac{1}{\lambda_{\phi}} \gtrsim 1 \; .
\end{equation}
The borderline values $\beta \simeq 1$ are of major interest because in that case one ends up with sizeable GWs as we show below. 

While the Universe is cooling down, the mass term of the field $\chi$ becomes more and more relevant, 
and the symmetry gets eventually restored. At this point the DW network dissolves. This can happen 
earlier, when the temperature $T$ drops below $m_{\phi}$, i.e., $m_{\phi} \gtrsim T$, 
so that the particles $\phi$ become non-relativistic, and their number density as well as variance $\langle |\phi|^2 \rangle$ experience an exponential falloff due to the appearance of the Boltzmann suppression factor. In what follows we mainly assume the limit of small $M_{\chi}$ and $m_{\phi}$, and comment on the situations where this approximation becomes invalid. We also discuss implications of small but non-zero $M_{\chi}$ for dark matter in the end of this section.

In the case $M_{\chi}=0$ and $m_{\phi}=0$, results of the previous sections are applicable. We put in contact the model described above with our numerical findings regarding DW evolution and GW spectrum. We mainly refer to the situation where we have observed the scaling regime for the 
DW network. However, as it has been discussed in Section~\ref{sec:gwnumerics}, GW spectra produced by melting DWs essentially coincide below a particular frequency independently of the scaling property. First, 
we express the constant $\alpha$ defined in Eq.~\eqref{alphadef} in terms of model constants:
\begin{equation}
\label{var}
\alpha=\sqrt{\frac{{\cal N} g^2}{12\lambda_{\chi}}} \; .
\end{equation}
Using the latter, we write for the Universe temperature at which the scaling regime is attained ~\eqref{initscaling} as follows: 
\begin{equation}
\label{scalingmodeltemp}
T_{sc} \simeq \frac{100~\mbox{MeV} \, \sqrt{{\cal N}}}{\sqrt{g_* (T_{sc})}} \cdot \left(\frac{g}{10^{-18}} \right)  \; .
\end{equation}
Here we also used the value $\tau_i/\tau_{sc} \simeq 1/25$ obtained numerically. Substituting Eq.~\eqref{scalingmodeltemp} into Eq.~\eqref{frtemp}, we obtain for the peak frequency:
\begin{equation}
\label{fpeakmodel}
f_{peak}  \simeq \frac{15 \, \mbox{nHz} \, \sqrt{{\cal N}}}{g^{1/3}_* (T_{sc})} \cdot \left(\frac{g}{10^{-18}} \right)  \; .
\end{equation}
Finally, in the model at hand the numerically obtained peak energy density of GWs~\eqref{peaknumerics} can be phrased as follows using~Eq.~\eqref{OmegaOmega}:
\begin{equation}
\label{peakone}
\Omega_{gw, peak} h^2_0 \simeq \frac{5 \cdot 10^{-11} \, {\cal N}^4}{g^{7/3}_* (T_{sc}) \, \beta^2} \;,
\end{equation}
where we also substituted the value $\xi_{sc} \approx 1.4$ derived from numerical simulations of melting DWs. 

The low frequency spectral shape of GWs from melting DWs is in a remarkable agreement with the recent PTA measurements of the GW background. 
This observation has been made in Ref.~\cite{Babichev:2023pbf} using the value $n=2$ originally obtained in Ref.~\cite{Babichev:2021uvl}. 
The updated spectral index $n \approx 1.6$ does not alter this conclusion, as the value measured by the PTAs reads $n =1.8 \pm 0.6$ at $68\%$ CL. 
Below we revisit the model constants needed to match also the observed amplitude of the signal. 

We require that the peak of GWs emitted by melting DWs corresponds to the maximal signal seen in PTAs estimated as 
$\Omega_{gw, peak} \sim 10^{-8}$. To match the predicted signal to this value, one chooses the constant $\beta$ to be as small as possible, i.e., close to its lower bound $\beta \simeq 1$. Furthermore, we choose ${\cal N}$ large enough, i.e., ${\cal N}=24$, to balance the suppression by $g^{7/3}_* (T_{sc})$; we assume $g_* (T_{sc}) \approx 40$. In particular, this can be achieved, if the field $\phi$ is in the adjoint representation of $\mbox{SU} (5)$ group. With these values one gets the peak value $\Omega_{gw, peak} h^2_0 \simeq 3 \cdot 10^{-9}$, which is somewhat below the maximum value observed by PTAs, albeit not dramatically, and a small variation 
of the scenario may improve the situation. One such variation will be considered shortly.  
The above assumption that the peak of predicted GWs corresponds to the maximal PTA signal, fixes $f_{peak} \simeq 32~\mbox{nHz}$. By virtue of Eq.~\eqref{fpeakmodel}, this value of $f_{peak}$ gets converted into a very small constant $g \simeq 1.5 \cdot 10^{-18}$. Together with $\beta \simeq 1$ this yields the self-interaction constant $\lambda_{\chi} \simeq 5 \cdot 10^{-72}$. Such tiny $\lambda_{\chi}$ are characteristic for axion-like particles, albeit we deal with a completely different scenario with different underlying symmetries.

Using the peak frequency value $f_{peak} \simeq 32~\mbox{nHz}$ and Eq.~\eqref{frtemp}, one can also extract the Universe temperature at the onset of scaling:
\begin{equation}
T_{sc} \simeq 120~\mbox{MeV} \; .
\end{equation}
This leads to quite severe constraints on the properties of the thermal field $\phi$. Namely, its mass should be well below $T_{sc}$. Otherwise, the field $\phi$ variance would experience Boltzmann suppression compared to Eq.~\eqref{var}. In this situation, the DW tension would be also exponentially suppressed compared to Eq.~\eqref{meltingtension}; no observable GWs are possible in this case. In fact, if we want to explain the NANOGrav data spanning one order of magnitude in the frequency range, a harsher condition is required $m_{\phi} \lesssim T_{sc}/10$. On the other hand, the mass $m_{\phi}$ must be above the BBN temperature $T_{BBN} \sim 1~\mbox{MeV}$, in order to avoid the conflict with the existing bound on the effective number of neutrino species~\cite{Planck:2018vyg}. Therefore, the mass $m_{\phi}$ is confined to a rather narrow range:
\begin{equation}
1~\mbox{MeV} \lesssim m_{\phi} \lesssim 10~\mbox{MeV} \; .
\end{equation}
Hence, the scenario at hand suggests existence of new particles at MeV scale.

In view of the shortcomings of the above scenario, let us consider its variation. Before we have considered the situation where the particles $\phi$ coexist with beyond the SM species in the primordial plasma. Now let us discuss the situation with non-standard early-time cosmology, where relativistic particles $\phi$ dominate the early Universe well before the BBN, while the number densities of SM species are negligible. We assume feeble couplings between the particles $\phi$ and the SM species triggering the decay of particles $\phi$ at some point in the early Universe and leading to the primordial bath populated by SM particles. As it follows, the number of relativistic degrees of freedom prior to the decay of particles $\phi$ reads $g_* (T)={\cal N}$ at $t \lesssim t_{dec}$. After that the number of relativistic degrees of freedom evolves in the standard fashion. Consequently, the fractional spectral energy density of GWs at peak reads
\begin{equation}
\Omega_{gw, peak} h^2_0 \simeq \frac{ 5 \cdot 10^{-11} \, {\cal N}^2}{g^{1/3}_* (T_{SM, dec}) \, \beta^2}   \; ,
\end{equation}
where $T_{SM, dec}$ is the temperature of the primordial plasma right after the decay in the SM species. We assume that the decay is nearly instantaneous.  In particular, for ${\cal N}=24$, $\beta \simeq 1$, and $g_* (T_{SM, dec}) \simeq 10$, one gets $\Omega_{gw, peak} \simeq  10^{-8}$, which is in a good agreement with PTAs keeping in mind that we deal with considerable departures from a power-law. 

The decay rate of relativistic particles $\phi$ is estimated as 
\begin{equation}
\Gamma_{\phi} \sim \frac{m_{\phi}}{3\,T_{dec}} \cdot \Gamma^0_{\phi} \; ,
\end{equation}
where $\Gamma_{\phi,0}$ is the decay rate of $\phi$-particles at rest. One can assume that the interaction of particles $\phi$ with the SM species is through some non-renormalisable operator suppressed by the scale $\Lambda$, so that the decay rate $\Gamma_{\phi,0}$ is estimated as 
\begin{equation}
\Gamma^0_{\phi} \sim \frac{m^3_{\phi}}{8\pi \Lambda^2} \; .
\end{equation}
The reheating happens when the Hubble parameter drops to the decay rate, $H\sim \Gamma_\phi$. 
Consequently, one can estimate the mass $m_{\phi}$ as follows: 
\begin{equation}
m_{\phi} \sim 2 \, {\cal N}^{1/8} \cdot \frac{\Lambda^{1/2} \cdot T^{3/4}_{dec}}{M^{1/4}_{Pl}} \; .
\end{equation}
Given that $m_{\phi} \lesssim T_{dec}$, because we consider the decay of relativistic particles, and that $T_{dec} \lesssim 10~\mbox{MeV}$, in order to explain the PTA signal, one gets the constraint on $\Lambda$: 
\begin{equation}
\Lambda \lesssim 2 \cdot 10^{7}~\mbox{GeV} \; .
\end{equation}
It is worth stressing that in this variation of the scenario we do not need to impose restrictive bounds on the particles $\phi$: they only need to be sufficiently light $m_{\phi} \lesssim 10~\mbox{MeV}$. On the other hand, the combination of parameters $m_{\phi}$ and $\Lambda$ is quite constrained. Indeed, besides the bound $T_{dec} \lesssim 10~\mbox{MeV}$, one should also fulfill the constraint $T_{dec} \gtrsim 4~\mbox{MeV}$ in order to do not spoil BBN~\cite{Hannestad:2004px}. Hence, one obtains
\begin{equation}
5~\mbox{MeV} \cdot \sqrt{\frac{\Lambda}{2 \cdot 10^{7}~\mbox{GeV}}} \lesssim m_{\phi} \lesssim 10~\mbox{MeV} \cdot \sqrt{\frac{\Lambda}{2 \cdot 10^{7}~\mbox{GeV}}} \; .
\end{equation}
We postpone further discussion of this scenario and embedding in a proper particle physics setup for the future. 

Finally, let us comment on the case, when the mass $M_{\chi}$ is non-zero. In this case, the field $\chi$ may be considered for the role of dark matter. There are two ways for dark matter production 
discussed in Ref.~\cite{Babichev:2021uvl}. Dark matter can be manifested via oscillations of the scalar $\chi$ around its minima following the direct second order phase transition, when DWs are formed. In this case, the mass $M_{\chi}$ is estimated as
\begin{equation}
\label{massfirst}
M_{\chi} \simeq 1.3 \cdot 10^{-16}~\mbox{eV} \, g^{4/3}_* (T_{sc})  \, \beta  \cdot \frac{1}{{\cal N}^2} \cdot
\frac{f_{peak}}{30~\mbox{nHz}}  \; .
\end{equation}
Here we ignored particle production by melting DWs, cf. Ref.~\cite{Kawasaki:2014sqa}. This will be considered elsewhere, but we expect the resulting mass to be in the ballpark of Eq.~\eqref{massfirst}. If the particles produced during the second order phase transition or emitted by DWs have been dissolved in the plasma, one can still produce DM via the inverse phase transition, which happens when the temperature drops below the mass $m_{\phi}$ and the expectation value $\eta$ turns into zero. 
As it has been discussed above, this happens only slightly before BBN. Masses of dark matter produced at the inverse phase transition~\cite{Ramazanov:2021eya, Babichev:2021uvl} (see also Refs.~\cite{Babichev:2020xeg, Ramazanov:2020ajq}) are fixed to be
\begin{equation}
\label{masssecond}
M_{\chi} \simeq \frac{0.9 \cdot 10^{-12}~\mbox{eV} \, g^{3/5}_* (T_{sc})\, \beta^{3/10}}{{\cal N}^{3/5}} \cdot \left(\frac{f_{peak}}{30~\mbox{nHz}} \right)^{6/5}  \cdot \sqrt{\frac{m_{\phi}}{10~\mbox{MeV}}}  \; .
\end{equation}
The masses~\eqref{massfirst} and~\eqref{masssecond} are updated to include a renewed estimate of the peak frequency in terms of the constant $g$. While it is not clear, which mass one should pick, apparently it is in the ultra-light range, such that Kerr black holes in the astrophysical range will be affected~\cite{Zeldovich1, Zeldovich2, Starobinsky:1973aij, Arvanitaki:2009fg}. It allows one to test the dark matter option of the suggested model of GW production by melting DWs.

\section{Discussion}
\label{sec:discussions}

Using numerical and analytic methods, we investigated in details melting DWs characterised by the time-decreasing tension.
It has been demonstrated that the network of melting DWs typically enters the scaling regime after some time likely defined by the ratio of wall width and the horizon size. Nevertheless, occurrence of scaling is not universal, as we have observed violation of the scaling law  for a large amplitude of initial scalar fluctuations relative to the vacuum expectation value. At the moment it is unclear whether this violation is inherent to the melting DW network or rather to the numerical procedure used. The latter may fail to distinguish walls from large-amplitude scalar waves at small scales, where the violation is mainly observed. Furthermore, we have not included a potential dissipation of large scalar fluctuations in the primordial plasma.

We also performed numerical simulations of GWs produced by the network of melting DWs. Compared to expectations of Refs.~\cite{Babichev:2023pbf, Babichev:2021uvl, Ramazanov:2023eau} where the spectral index $n=2$ in the low-frequency range has been quoted, we rather obtain the spectral index $n \approx 1.6$. We investigated the possible origin of this departure: 
while the result $n=2$ corresponds to the ``idealised" situation with the everlasting source, the finite duration of the source inevitably leads to the tilt relative to $n=2$. Note that the prediction $n \approx 1.6$ is also in a very good agreement with the recent PTA findings giving $n =1.8 \pm 0.6$ at $68\%$ CL. It is crucial that the spectral index $n \approx 1.6$ holds whether the scaling law is obeyed or not. Therefore, we expect that the future PTA measurements of the spectral index $n$ will be able to conclude whether the nHz GW background is explained by melting DW emission. Note also that phenomenology of melting DWs is not limited to nHz GWs. Depending on the constants in the underlying model (e.g., the constant $g$ in the model of Section~\ref{sec:applications}), one can get GWs in the frequency range covering all current and planned GW experiments, e.g., LISA~\cite{LISA:2017pwj}, TianQin~\cite{TianQin:2015yph}, or Einstein Telescope~\cite{Hild:2010id}. 

Results of this work can be also important from the viewpoint of differentiating GWs produced by melting DWs and other possible sources. Clearly, the spectral shape discussed in this work is distinct from most other sources such as first order phase transitions, constant tension DWs, and (stable) cosmic strings~\cite{Caprini:2018mtu}. However, metastable cosmic strings~\cite{Vilenkin:1982hm, Preskill:1992ck, Monin:2008mp} are capable of producing GWs with the spectral shape $n =2$, which exhibited degeneracy with the value in the case of melting DWs as predicted in Refs.~\cite{Buchmuller:2023aus}. The present analysis lifts this degeneracy enabling to distinguish metastable cosmic strings and melting DWs in the current and future measurements of the GW background.  

\section*{Acknowledgments} 
Lattice simulations have been carried out on the computer cluster of the Theoretical Division of INR RAS and on the cluster ``Phoebe" at CEICO (FZU). We are grateful to Josef Dvo\v r\'a\v cek for the help in using the ``Phoebe" cluster. The work of D.~G. and I.~D. was supported by the scientific program of the National Center for Physics and Mathematics, section 5 ``Particle Physics and Cosmology", stage 2023-2025. The work of E.~B. was supported by ANR grant StronG (ANR-22-CE31-0015-01). A.~V. acknowledges the support from the Czech Science Foundation, GA\v{C}R, project number 24-13079S.

\section*{Appendix. Analytical estimation of gravitational wave spectrum}

In this Appendix, we estimate the function $A(u_{sc}, u)$ defined in Eq.~\eqref{A}, where $u_{sc} \equiv k\tau_{sc}$ and $u \equiv k\tau$, which is crucial for getting GW spectrum in the interesting frequency range. For this purpose, we make a 
simplifying assumption that the reduced power spectrum ${\cal P} (u', u'')$ is totally uncorrelated, i.e., it has the form~\cite{Caprini:2009fx}
\begin{equation}
\label{simple}
{\cal P} (u', u'')= {\cal J} \cdot {\cal F} (u') \cdot \delta (u'-u'') \; ,
\end{equation}
where ${\cal F} (u)$ and ${\cal J}$ are the function and the constant to be defined below. 
To understand the behaviour of the function ${\cal F} (u)$, let us make the following trick. We consider only GWs emitted by the source in a short time interval $\tau_2-\tau_1 \ll 1/k$, where 
$\tau_{sc} \ll \tau_1<\tau_2 \ll \tau$. The spectral energy density of GWs produced in this time interval is estimated as 
\begin{equation}
\label{simplespectrum}
\frac{d\rho_{gw} (\tau)}{d \ln k} \simeq 64 \pi^2 \, G_N \, \rho^2_{wall} (\tau_1) \,
a^2 (\tau_1) \, k^2 \, \tau^4_1 \cdot \left(\frac{a(\tau_1)}{a(\tau)} \right)^4 \cdot \frac{\tau_2-\tau_1}{\tau_1} \cdot {\cal J} \cdot {\cal F} (u_1) \; .
\end{equation}
Here we used Eqs.~\eqref{spectrum} and~\eqref{simple}, and we also assumed that $k\tau \gg 1$. Furthermore, we assumed that the function ${\cal F} (u)$ has approximately the same shape in the scenarios of constant tension and melting DWs, where GWs are mainly sourced by long DWs (see also below). According to Eq.~\eqref{simplespectrum}, this is equivalent to saying that the spectrum of GWs generated in the interval $\tau_2-\tau_1$ is approximately the same for two types of DWs. In favour of this statement let us notice that melting DW tension can be considered constant for a very short interval $\tau_2-\tau_1$ assumed. Thus we can use results of numerical simulations performed  for constant tension walls~\cite{Hiramatsu:2013qaa, Dankovsky:2024zvs}. For $u_1 \simeq u_2  \ll 2\pi$, one has ${\cal F} (u_1) \propto u_1 \propto k$, 
because only in that case one can reproduce the correct behaviour $d\rho_{gw}/d\ln k \propto k^3$ following from causality considerations~\cite{Caprini:2009fx}. In the regime, $u_1 \gg 2\pi$, the GW spectrum of constant tension walls experiences the power-law falloff $d\rho_{gw}/d\ln k \propto 1/k^{q}$, where the exponent $q$ is fixed by numerical simulations to be $q \approx 1.5$~\cite{Dankovsky:2024zvs}. This behaviour is reproduced, provided the function ${\cal F} (u_1)\propto 1/u^{q+2}_1$, i.e., ${\cal F} (u_1) \propto 1/u^{3.5}_1$. 
Interpolating the behaviour from $u_1 \ll 2\pi$ and high $u_1 \gg 2\pi$ to $u_1 \simeq 2\pi$, we consider the function ${\cal F} (u_1)$ of the form:
\begin{equation}
\label{simpletwo}
{\cal F} (u_1) \sim
\begin{cases}
\frac{u_1}{2\pi} \qquad \qquad ~ u_1 \leq 2\pi\\ 
\left(\frac{2\pi}{u_1} \right)^{3.5} \qquad u_1> 2\pi \; .
\end{cases}
\end{equation}
As it follows from Eqs.~\eqref{A} and~\eqref{simple},  
\begin{equation}
A (u_{sc}, u)=\int^{u}_{u_{sc}} \frac{du_1}{u_1} {\cal F} (u_1) \; .
\end{equation}
From Eq.~\eqref{simpletwo} we get for the most interesting $u$ and $u_{sc}$: 
\begin{equation}
\label{simplethree}
A (u_{sc}, u) \sim {\cal J} \cdot
\left( 1.3-\frac{u_{sc}}{2\pi}-0.3\cdot \left(\frac{2\pi}{u} \right)^{3.5} \right)  \qquad u \gg 2\pi \quad u_{sc} \ll 2\pi\; .
\end{equation}
Taking the limits $u_{sc} \rightarrow 0$ and $u \rightarrow \infty$, we obtain $A(0, \infty) \sim 1.3 \; {\cal J}$. On the other hand, 
for generic $u \gg 2\pi$ and $u_{sc} \ll 2\pi$, one arrives at the $k$-dependence in the brackets in Eq.~\eqref{estimate}. In order to define the peak energy density of GWs, one should also consider the regime $u_{sc}>2\pi$, in which case one gets 
\begin{equation}
\label{a}
A (u_{sc}, u) \sim 0.3 \cdot {\cal J} \cdot \left(\frac{2\pi}{u_{sc}} \right)^{3.5} \qquad u \gg u_{sc} >2\pi \; .
\end{equation}
In principle, one may apply this estimate also for ultraviolet modes with $u_{sc} \gg 1$. However, in this regime the result can be sensitive to abundance of small closed DWs, and the scenarios with melting and constant tension DWs are markedly different in this regard, as it is discussed in Section~\ref{sec:dwevolution}.

Now let us determine the constant ${\cal J}$. This can be obtained from Eq.~\eqref{scalingdef}, where we contract both sides with polarisation tensors, perform the Fourier transformation with momenta ${\bf k}$ and ${\bf q}$, divide the result by $\tau''$, integrate over $\tau''$ and take the limit of equal space points:
\begin{equation}
\label{tricky}
\int^{\infty}_0 \frac{d\tau''}{\tau''} \cdot \langle \Pi_{ij} ({\bf x}, \tau') \Pi_{ij} ({\bf x}, \tau'') \rangle=16 \pi \int^{\infty}_0 \frac{d\tau''}{\tau''} \cdot \rho_{wall} (\tau') \rho_{wall} (\tau'') (\tau' \tau'')^{3/2} 
\int^{\infty}_0 dk k^2 {\cal P} (k\tau', k\tau'') \; .
\end{equation}
Again inserting Eq.~\eqref{simple} and $\delta (u'-u'')=\delta(\tau'-\tau'')/k$, integrating the r.h.s. of Eq.~\eqref{tricky} over $\tau''$, we obtain 
\begin{equation}
\label{start}
\int^{\infty}_0 \frac{d\tau''}{\tau''} \cdot \langle \Pi_{ij} (0, \tau') \Pi_{ij} (0, \tau'') \rangle=16 \pi \rho^2_{wall} (\tau') \, {\cal J} \, 
\int^{\infty}_0 du' u' {\cal F} (u') \; ,
\end{equation}
where we got rid of ${\bf x}$-dependence of the correlator on the l.h.s. by making use of statistical homogeneity. 
We integrate over $u'$ using Eq.~\eqref{simpletwo} and obtain 
\begin{equation}
\label{int}
\int^{\infty}_0 du' u' {\cal F} (u') \simeq 4\pi^2 \; .
\end{equation}
The correlator on the l.h.s. of Eq.~\eqref{start} can be rewritten as follows: 
\begin{equation}
\label{intermede}
\begin{split}
&\langle \Pi_{ij} (0, \tau') \Pi_{ij} (0, \tau'') \rangle =\frac{1}{(2\pi)^6 a^2 (\tau') a^2 (\tau'')} \cdot 
\int d{\bf y} d{\bf y}' \langle T_{kl} ({\bf y}, \tau'), T_{k'l'} ({\bf y}', \tau'') \rangle \times \\ \times & \int d{\bf k} d{\bf k}' e^{-i{\bf k} {\bf y} -i{\bf k}' {\bf y}'} \cdot \Lambda_{ij, kl} ({\bf \hat{k}}) \Lambda_{ij, k'l'} ({\bf \hat{k}}') \; .
\end{split}
\end{equation}
We again use the property of statistical homogeneity and write 
\begin{equation}
\langle T_{kl} ({\bf y}, \tau'), T_{k'l'} ({\bf y}', \tau'') \rangle =
\langle T_{kl} ({\bf z}, \tau'), T_{k'l'} (0, \tau'') \rangle \; ,
\end{equation}
and we replace ${\bf y}$ by the new variable ${\bf z}={\bf y}-{\bf y}'$. Thus, we can rewrite Eq.~\eqref{intermede} as 
\begin{equation}
\begin{split}
&\langle \Pi_{ij} (0, \tau') \Pi_{ij} (0, \tau'') \rangle =\frac{1}{(2\pi)^6 a^2 (\tau') a^2 (\tau'')} \cdot 
\int d{\bf z} d{\bf y}' \langle T_{kl} ({\bf z}, \tau'), T_{k'l'} (0, \tau'') \rangle \times \\ \times & \int d{\bf k} d{\bf k}' e^{-i{\bf k} {\bf z}-i{\bf k}{\bf y}' -i{\bf k}' {\bf y}'} \cdot \Lambda_{ij, kl} ({\bf \hat{k}}) \Lambda_{ij, k'l'} ({\bf \hat{k}}') \; .
\end{split}
\end{equation}
The integrals over ${\bf y}'$ and ${\bf k'}$ are easily taken and one gets 
\begin{equation}
\label{interme}
\begin{split}
&\langle \Pi_{ij} (0, \tau') \Pi_{ij} (0, \tau'') \rangle =\frac{1}{(2\pi)^3  a^2 (\tau') a^2(\tau'')} \int d{\bf z}  
\langle T_{kl} ({\bf z}, \tau') T_{k'l'} (0, \tau'') \rangle \times \\ &\times \int d{\bf k} e^{-i {\bf k z}} \Lambda_{ij, kl} ({\bf \hat{k}}) \Lambda_{ij, k'l'} (-{\bf \hat{k}})\; .
\end{split}
\end{equation}
Using the properties of the $\Lambda$-tensor $\Lambda_{ij, k'l'}(-{\bf \hat{k}})=\Lambda_{ij, k'l'} ({\bf \hat{k}})$ and $\Lambda_{ij,kl} ({\bf \hat{k}}) \Lambda_{ij, k'l'} ({\bf \hat{k}})=\Lambda_{kl, k'l'} ({\bf \hat{k}})$, one obtains
\begin{equation}
\label{interme}
\begin{split}
&\langle \Pi_{ij} (0, \tau') \Pi_{ij} (0, \tau'') \rangle =\frac{1}{(2\pi)^3  a^2 (\tau') a^2(\tau'')} \int d{\bf z}  
\langle T_{kl} ({\bf z}, \tau') T_{k'l'} (0, \tau'') \rangle \times \\ &\times \int d{\bf k} e^{-i {\bf k z}} \Lambda_{kl, k'l'} ({\bf \hat{k}}) \; .
\end{split}
\end{equation}
In the next step, we loosely replace $\Lambda$-tensor in Eq.~\eqref{interme} by its average over directions, i.e., 
\begin{equation}
\label{lambdareplace}
\Lambda_{kl, k'l'} ({\bf \hat{k}}) \sim \bar{\Lambda}_{kl, k'l'} = \frac{1}{4\pi} \int d{\bf \hat{k}} \Lambda_{kl, k'l'} ({\bf \hat{k}}) =\frac{1}{30} 
\cdot \left(11 \delta_{kk'} \delta_{ll'} -4 \delta_{kl} \delta_{k'l'} +\delta_{kl'} \delta_{k'l} \right) .
\end{equation}
Say it another way, we assume that dropping the contribution due to the difference $\delta \Lambda_{kl, k'l'}=\Lambda_{kl, k'l'}-\bar{\Lambda}_{kl, k'l'}$ does not change the final estimate considerably. 
With the replacement~\eqref{lambdareplace}, one can easily take the integral over the momentum ${\bf k}$ in Eq.~\eqref{interme}, and get
\begin{equation}
\label{1one}
\begin{split}
&\int \frac{d\tau''}{\tau''} \langle \Pi_{ij} (0, \tau') \Pi_{ij} (0, \tau'') \rangle \sim \frac{2}{15a^2 (\tau') } \times \\ \times &\int \frac{d\tau''}{a^2 (\tau'') \tau''}
\cdot \left[ 3\langle T_{kl} (0, \tau') T_{kl} (0, \tau'') \rangle -\langle T_{kk} (0, \tau') T_{ll} (0, \tau'') \rangle \right] \; , 
\end{split}
\end{equation}
where we have integrated both parts over $\ln \tau''$. We make the estimate
\begin{equation}
\int \frac{d\tau''}{a^2 (\tau'') \tau''} \left[ 3\langle T_{kl} (0, \tau') T_{kl} (0, \tau'') \rangle -\langle T_{kk} (0, \tau') T_{ll} (0, \tau'') \rangle\right]  \sim \frac{1}{a^2 (\tau')} \cdot \left[3T^2_{kl} (0, \tau')-T^2_{kk} (0, \tau') \right]\; .
\end{equation}
This assumes that realistically the integral over $\ln \tau''$ is saturated in the interval corresponding to the Hubble time, i.e., $\Delta \ln \tau'' \sim 1$, and that components of the stress-energy tensor are not changing much in this interval. We also omitted averaging over realisations. Let us consider for simplicity a sufficiently flat DW perpendicular to the $z$-axis at the time $\tau'$ with no loss of generality. Recall that such a configuration is close to the realistic one, if the DW is in the scaling regime. For such a configuration one can show that 
\begin{equation}
3T^2_{kl} (0, \tau')-T^2_{kk} (0, \tau') =2 (\partial_z \chi)^4 \; .
\end{equation}
We compare this with the DW energy density: 
\begin{equation}
\label{two2}
\rho_{wall} \simeq \frac{1}{2a^2} (\partial_z \chi)^2 +V(\chi) \simeq \frac{1}{a^2} (\partial_z \chi)^2 \; . 
\end{equation}
Note that we have neglected corrections due to the DW velocity here, 
which are ${\cal O} (\frac{v^2}{c^2})$. These are not necessarily small corrections, but they can be omitted in view of the order of magnitude estimation. Combining Eqs.~\eqref{1one}-\eqref{two2}, we get
\begin{equation}
\int \frac{d\tau''}{\tau''} \langle \Pi_{ij} (0, \tau') \Pi_{ij} (0, \tau'')  \rangle \sim \frac{4\rho^2_{wall} (\tau')}{15} \; .
\end{equation}
Comparing this with Eq.~\eqref{start} and using Eq.~\eqref{int}, we obtain
\begin{equation}
\label{j}
{\cal J} \sim \frac{1}{240\pi^3} \; .
\end{equation}
Finally, substituting this estimate of ${\cal J}$ into Eq.~\eqref{simplethree} and the latter into Eq.~\eqref{spectrum} from the main text, neglecting oscillating terms, we end up with the estimate~\eqref{estimate}.

\end{document}